\documentclass[%
 aip,
 amsmath,amssymb,
reprint,%
]{revtex4-1}

\usepackage{graphicx}
\usepackage{dcolumn}
\usepackage{bm}
\usepackage{hyperref}
\usepackage[utf8]{inputenc}
\usepackage[T1]{fontenc}
\usepackage{mathptmx}
\usepackage{etoolbox}
\usepackage{mathtools}  
\usepackage{braket} 	
\usepackage{color} 		
\makeatletter
\def\@email#1#2{%
 \endgroup
 \patchcmd{\titleblock@produce}
  {\frontmatter@RRAPformat}
  {\frontmatter@RRAPformat{\produce@RRAP{*#1\href{mailto:#2}{#2}}}\frontmatter@RRAPformat}
  {}{}
}%
\makeatother

\newcommand{\HH}{{\mathcal{H}}}

\newcommand{\UU}{\textrm{U}}
\newcommand{\rf}{\textrm{rf}}
\newcommand{\dd}{\textrm{d}}

\newcommand{\rr}{\textrm{r}}
\newcommand{\iso}{\textrm{iso}}

\newcommand{\II}{\textrm{I}}

\newcommand{\Fou}{\mathcal{F}}
\newcommand{\iFou}{\mathcal{F}^{-1}}

\newcommand{\SSS}{\textrm{S}}
\newcommand{\tr}{\textrm{tr}}

\graphicspath{{figures/}}
\begin{document}
\preprint{AIP/123-QED}

\title[Continuous Floquet Theory in NMR]{A continuous approach to Floquet theory\\ for pulse-sequence optimization in solid-state NMR}

\author{Matías Chávez}
\affiliation{Physical Chemistry, ETH Zürich, Vladimir-Prelog-Weg 2, 8093 Zürich, Switzerland}

\author{Matthias Ernst}%
\email{maer@ethz.ch}
\affiliation{Physical Chemistry, ETH Zürich, Vladimir-Prelog-Weg 2, 8093 Zürich, Switzerland}

\date{\today}

\begin{abstract}
We present a framework that uses a continuous frequency space to describe and design solid-state NMR experiments.   The approach is similar to the well established Floquet treatment for NMR, but is not restricted to periodic Hamiltonians and allows the design of experiments in a reverse fashion. The framework is based on perturbation theory on a continuous Fourier space, which leads to effective, i.e., time-independent, Hamiltonians. It allows the back calculation of the pulse scheme from the desired effective Hamiltonian as a function of spin-system parameters. We show as an example how to back calculate the rf irradiation in the MIRROR experiment from the desired chemical-shift offset behaviour of the sequence.
\end{abstract}
\maketitle
%
%
\section{Introduction}
Time-dependent Hamiltonians are very common in (solid-state) NMR due to sample rotation, e.g., magic angle spinning (MAS) \cite{Andrew:1958we,Andrew:1959uy,Lowe:1959ur} or even rotations about multiple axis as implemented in double rotation (DOR) \cite{Samoson:1988fm} or dynamic-angle spinning (DAS) \cite{Mueller:1990wf}, and the application of pulse sequences that can be described by interaction-frame transformations.\cite{Haeberlen:1976uz,Mehring:1983wm} Calculating the time evolution of the density operator under a time-dependent Hamiltonian is usually done numerically using time-slicing methods where we assume that the Hamiltonian is time constant for small enough time periods, $\tau$. However, in many cases an analytical solution to the Liouville-von Neumann equation is advantageous especially if predictions of the response of the sequence on spin-system parameters are of interest. Since the late 1960s, various methods have been developed to approximate such time-dependent Hamiltonians using time-independent representations. This can be achieved using average Hamiltonian theory (AHT), \cite{Haeberlen:1968wu,Haeberlen:1976uz,Maricq:1982wf,Ernst:1990vd} Floquet theory \cite{Floquet:1883vq,Shirley:1965tn,Scholz:2010hq,Leskes:2010dx,Ivanov:2021uy} or other less-established methods like the Fer expansion \cite{MADHU:2006fx,Takegoshi:2015bd}, the Floquet-Magnus expansion \cite{Mananga:2016hh,Mananga:2018jn}, or the path-sum method. \cite{Giscard:2020dz}
\\
Average Hamiltonian theory \cite{Haeberlen:1968wu,Haeberlen:1976uz,Maricq:1982wf,Ernst:1990vd} can be used for periodic Hamiltonians and is based on the Magnus expansion. \cite{Blanes:2009co} It generates a series of  time-independent average Hamiltonians that approximate the time evolution of the density operator over the basic time period with increasing accuracy. Limitations of AHT are the stroboscopic observations at integer multiples of the basic time period, e.g., the difficulty to describe spinning side bands in MAS spectra, and the difficulty to describe Hamiltonians with multiple incommensurate frequencies that appear for example in homonuclear \cite{Mote:2016cr} or heteronuclear \cite{Ernst:2003hx} decoupling experiments. Despite these limitations, AHT is one of the most used techniques in NMR to develop and optimize pulse sequences.
\\
In Floquet theory \cite{Floquet:1883vq,Shirley:1965tn,Scholz:2010hq,Leskes:2010dx,Ivanov:2021uy} the finite-dimensional time-dependent spin-Hilbert space Hamiltonian is replaces by an infinite-dimensional but time-independent Floquet Hamiltonian. Both representations are equivalent and describe the identical time evolution of the density operator. Floquet theory can be used to describe periodic time-dependent Hamiltonians without the requirement of stroboscopic sampling, i.e., side bands under MAS are predicted correctly \cite{Ivanov:2021uy} and also multiple incommensurate time dependencies can be included using multi-mode Floquet theory. \cite{Scholz:2010hq,Leskes:2010dx,Ivanov:2021uy} Using the time-independent Floquet Hamiltonian makes no approximations but it is not always easy to get physical insights from the infinite-dimensional matrices. Therefore, operator-based Floquet theory \cite{Augustine:1995uf,Boender:1996wu,Boender:1998vz,Ernst:2005ic,Ramachandran:2005bi,Scholz:2007bs}, that is based on the van Vleck perturbation treatment \cite{VanVleck:1929fp,Primas:1961vi,Primas:1963tg} in the Floquet space was developed. Such a treatment allows the analytical block diagonalization of the Floquet Hamiltonian and a subsequent projection back into the spin-Hilbert space, generating a series of effective Hamiltonians that describe the time evolution of the density operator. Floquet theory and especially the operator-based implementation has become an important tool for the understanding of magnetic-resonance experiments and also for the design of new experiments.\cite{Ivanov:2021uy}
\\
In this publication, we present a modified version of Floquet theory that is based on a continuous frequency space and not a discrete Fourier series. We show that the modified approach allows the description of sequences of limited length and also allows, in some cases, a simpler back calculation of pulse-sequence parameters. 
%
%
\section{Theory}
%
%
\subsection{Standard Floquet theory}
The formalism presented in this article is similar to the formulation of Floquet theory used in magnetic resonance. Therefore we first want to give a short review of the Floquet treatment and subsequently present the new formalism. A more detailed derivation can be found in the literature. \cite{Shirley:1965tn,Leskes:2010dx,Ivanov:2021uy}
The time dependence of the density operator is given by the Liouville-von Neumann equation
\begin{align}
\dot{\sigma}(t) = -i\left[\HH(t), \sigma(t)\right]
\label{eq:LvN}
\end{align}
with the solution
\begin{align}
\sigma(t) = \UU(t) \sigma(0) \UU^{-1}(t)
\label{eq:prop}
\end{align}
where the propagator is formally given by
\begin{align}
\UU(t) = {\mathcal{T}}\exp\left(-i\int\limits_0^t \HH(\tau)d\tau\right)
\end{align}
and $\mathcal{T}$ is the Dyson time-ordering operator. \cite{Dyson:1949tf}
An equivalent formulation and a good starting point for the derivation of Floquet theory is the differential equation in the propagator
\begin{align}
\dot{\UU}(t) = -i\,\HH(t) \UU(t)
\label{eq:SG}
\end{align}
Floquet theory for magnetic resonance is based on the assumption that the Hamiltonian is periodic and, therefore, the Schrödinger equation is a periodic differential equation.
We can incorporate the periodicity of the Hamiltonian explicitly by expanding it in a Fourier series as
\begin{align}
\HH(t) = \sum_k \HH^{(k)}e^{i k \omega t}
\label{eq:H_FS}
\end{align}
The form of the solution of the periodic differential Eq. (\ref{eq:SG}) is given by the Floquet theorem \cite{Floquet:1883vq} as
\begin{align}
\UU(t) = u(t)e^{i \Lambda t}u^{-1}(0)
\label{eq:U_Floquet}
\end{align}
where $\Lambda$ is a diagonal real matrix because $\HH(t)$ is a Hermitian operator. The matrix $u(t)$ has the same periodicity as the Hamiltonian and can also be expressed as a Fourier series
\begin{align}
u(t) = \sum_{n} u_n e^{i n\omega t}
\label{eq:u_FS}
\end{align}
Notice that now the time dependence of the operators is only in the exponent and, therefore, inserting Eqs. (\ref{eq:H_FS}) - (\ref{eq:u_FS}) into Eq. (\ref{eq:SG}) will lead to the algebraic equation
\begin{align}
\sum_{k} (\HH^{(n-k)} +\omega \delta_{n k}) u_k  = \Lambda u_n
\label{eq:eigenvalue_equation_u}
\end{align}
This equation is equivalent to a time-independent Schrödinger equation for the Fourier coefficients $u_n$, where the Hamiltonian
\begin{align}
\HH_F = \sum_{k} \HH^{(n-k)} +\omega \delta_{n k}
\end{align}
is called the Floquet Hamiltonian. 
It is convenient to represent the Floquet Hamiltonian and $u_n$ in a product space  of the spin Hilbert space and the Fourier space, where each basis state corresponds to a Fourier harmonic. Typically the basis states are written as $\ket{n,\mu}=\ket{n}\otimes\ket{\mu}$, where $\ket{\mu}$ denote the basis states of the spin Hilbert space, and $\ket{n}$ are the basis states of the Fourier space. Throughout this article Greek letters are used for the spin basis and Latin letters for the Fourier harmonics. Notice, that there are infinitely many Fourier harmonics ($n\in \mathbb{Z}$), hence the matrix representation of the operators has infinite dimensions as well. The explicit representation of $\HH_F$ \cite{Scholz:2010hq,Leskes:2010dx,Ivanov:2021uy} is given by:
\begin{align}
\HH_F= \sum_{n} F_n\otimes\HH^{(n)} +\omega F_z\otimes \mathbf{1}
\label{eq:floquet_hamiltonian}
\end{align}
Here, the $F$ operators act on the Fourier space and are defined by $F_z\ket{n}=n\ket{n}$ and $F_n\ket{m}=\ket{m+n}$. The solution of  Eqs. (\ref{eq:eigenvalue_equation_u}) or (\ref{eq:floquet_hamiltonian}) leads to the formal solution of Eq. (\ref{eq:SG}) expressed in the spin Hilbert space:
\begin{align}
\UU(t) =  \sum_{n} \bra{n}\exp\left(i \HH_F t\right)\ket{0} e^{i n \omega t}
\end{align}
In conclusion, we solved the Liouville-von Neumann equation (Eq. (\ref{eq:LvN})) by  converting it to an algebraic equation using Fourier series expansions and the Floquet theorem assuming a periodic time-dependent Hamiltonian. As a basis we chose the product of the spin Hilbert space and Fourier space basis, which uses Fourier harmonics.
Typically van Vleck perturbation theory is applied on the Floquet Hamiltonian to obtain effective Hamiltonians. The procedure can be found in the literature \cite{Ramesh:2001uz,Vinogradov:2001ui,Ernst:2005ic,Scholz:2007bs, Leskes:2010dx, Ivanov:2021uy} and leads to the first and second-order Hamiltonian
\begin{align}
\bar{\HH}^{(1)} = \HH^{(0)}
\end{align}
and
\begin{align}
&\bar{\HH}^{(2)} = \frac{1}{2}
\sum\limits_{n\neq 0}
\frac{[\HH^{(n)}
	,\HH^{(-n)}]}
{n \omega }
\end{align}
Extending operator-based Floquet theory to multiple frequencies is straightforward. The details can be found in several reviews. \cite{Scholz:2010hq,Leskes:2010dx,Ivanov:2021uy} 
%
%
\subsection{Frequency-domain formulation}
Let us consider an arbitrary time-dependent Hamiltonian that is not necessarily periodic in time. In this case we cannot use the Floquet theorem nor a Fourier series to solve Eq. (\ref{eq:LvN}), since both require periodicity. However, we can use similar methods such as the Fourier transform to obtain an algebraic equation from Eq. (\ref{eq:SG}). Instead of a Fourier series expansion, we apply the Fourier transformations to each matrix element of the Hamiltonian $\HH(t)$ and the propagator $\UU(t)$. This will lead to a similar description, where the Fourier transformed operators take the role of the Fourier coefficients in the Floquet approach. In addition, Fourier transformation leads to a continuous spectrum of frequencies instead of the Fourier harmonics.\\
For the derivation of the generalized framework we again start from the differential equation Eq. (\ref{eq:SG}).
As a first step we use the Fourier transformation (element wise) to define a frequency-domain Hamiltonian and propagator as
\begin{align}
&\HH(t) = \iFou\{{\widehat{\HH}(\Omega)}\} = \tau \int\limits_{-\infty}^\infty\widehat{\HH}(\Omega)e^{i \Omega t}\dd\Omega\\
&\widehat{\HH}(\Omega) = \Fou\{{\HH(t)}\} = \frac{1}{2\,\pi\,\tau} \int\limits_{0}^\tau\HH(t)e^{-i \Omega t}\dd t\\
&\UU(t) = \iFou\{{\widehat{\UU}(\Omega)}\} = \tau \int\limits_{-\infty}^\infty\widehat{\UU}(\Omega)e^{i \Omega t}\dd\Omega\\
&\widehat{\UU}(\Omega) = \Fou\{\UU(t)\} = \frac{1}{2\,\pi\,\tau} \int\limits_{0}^\tau \UU(t) e^{-i \Omega t}\dd t
\end{align}
The Fourier transformation is normalized such that it leads to a correspondence with the standard Floquet approach. Thus, the presented theory is a generalization of standard Floquet theory and reproduces all of its results. Notice that the normalization also has to ensure $\Fou\{\iFou\{f(x)\}\}==f(x)$.
The 'wide' hats on $\widehat{\UU}(\Omega)$ and $\widehat{\HH}(\Omega)$ indicate that these are frequency-domain operators. The underlying frequency domain is continuous, orthonormal, and complete, hence conceptually similar to the position or momentum basis used, for example, to describe a free particle. 
Inserting the expressions in Eq. (\ref{eq:SG}) and using the convolution theorem leads to
\begin{align}
i\,\iFou \{\Omega\, \widehat{\UU}(\Omega)\}
&= -i\,\iFou \{ \widehat{\HH}(\Omega)\}\, \iFou \{ \widehat{\UU}(\Omega)\}\notag\\
&= -i\, \iFou\{\widehat{\HH}(\Omega) * \widehat{\UU}(\Omega)\}
\label{eq:conv}
\end{align}
where $*$ symbolizes the convolution operation in frequency space. 
Applying Fourier transformation to Eq. (\ref{eq:conv}) results in the time-independent equation 
\begin{align}
\Omega\, \widehat{\UU}(\Omega)
  = - \widehat{\HH}(\Omega) * \widehat{\UU}(\Omega)
\label{eq:convol}
\end{align}
Since convolution with the delta function is an identity operation we can write Eq. (\ref{eq:convol}) as 
\begin{align}
\Omega\, \delta(\Omega)*\widehat{\UU}(\Omega)
= - \widehat{\HH}(\Omega) * \widehat{\UU}(\Omega)
\end{align}
and express it as a single integral
\begin{align}
\int\limits_{-\infty}^\infty \big[\Omega \,\delta(\Omega-\Omega') + \widehat{\HH}(\Omega-\Omega')\big] \widehat{\UU}(\Omega')\, \dd\Omega' = 0
\end{align}
Since this has to be fulfilled for every possible $\widehat{\UU}(\Omega')$ the kernel has to vanish
\begin{align}
\big[\Omega \,\delta(\Omega-\Omega') + \widehat{\HH}(\Omega-\Omega')\big] \widehat{\UU}(\Omega') = 0
\end{align}
Notice, that this is an eigenvalue equation of $\widehat{\UU}(\Omega')$ with the eigenoperator  
\begin{align}
\HH_F =  \Omega \,\delta(\Omega-\Omega') + \widehat{\HH}(\Omega-\Omega')
\label{F}
\end{align}
Similar to the standard Floquet treatment, we can express this Hamiltonian in the orthonormal product basis $\ket{\Omega,\mu} = \ket{\Omega} \otimes \ket{\mu}$. As before the states $\ket{\mu}$ denote the basis states of the spin Hilbert space, but $\ket{\Omega}$ are the basis states of the \textit{continuous} Fourier space.
Notice that the operators $\widehat{\HH}(\Omega-\Omega')$ and $\Omega$ in $\HH_{F}$ live in different subspaces, i.e the spin and the frequency space, respectively.
Hence the Hamiltonian in the product basis takes the form 
\begin{align}
\bra{\Omega,\mu} \HH_{F} \ket{\Omega',\nu} = \Omega\, \delta(\Omega-\Omega')\,\otimes\delta_{\mu \nu} \notag\\
+\mathbf{1}\otimes  \widehat{\HH}_{\mu \nu}(\Omega-\Omega')
\label{F2}	
\end{align}
where $\widehat{\HH}_{\mu \nu}(\Omega) = \bra{\mu} \widehat{\HH}(\Omega) \ket{\nu}$ are the matrix elements of the spin-system Hamiltonian at frequency $\Omega$.

In full analogy to the Floquet description (Eq. \eqref{eq:floquet_hamiltonian}), $\HH_{F}$ can be formulated in a basis-free operator form as
\begin{align}
\HH_{F} =\int  \widehat{D}(\Omega) \otimes\widehat{\HH}(\Omega)\, \dd \Omega + \widehat{\Omega} \otimes \mathbf{1}
\label{Flo}
\end{align}
with the frequency operator
\begin{align}
\widehat{\Omega} \ket{\Omega} = \Omega \ket{\Omega} &\iff \widehat{\Omega} = \int
\Omega'  \ket{\Omega'}\bra{\Omega'}\, \dd \Omega'
\end{align}
and the frequency-shift operator
\begin{align}
&\widehat{D}(\Omega_0) \ket{\Omega} = \ket{\Omega + \Omega_0} \\
&\widehat{D}(\Omega_0) = \int_{\Omega'} \ket{\Omega'+\Omega_0}\bra{\Omega'} \dd \Omega'.
\end{align}
The equivalence of Eq. \eqref{Flo} and Eq. \eqref{F} is shown in the supplementary information (SI).
Notice that we introduced the frequency operator $\widehat{\Omega}$, which is similar to a position operator, but acting on the frequency space. It takes the place of the number operator $F_z$ in standard Floquet theory. The frequency-shift operator $\widehat{D}(\Omega)$ is conceptually similar to the usual translation operator in the position space. It takes the place of the generalized ladder operators $F_n$ in standard Floquet theory. Both of this operators act exclusively on the frequency domain.
\\
As in the standard Floquet treatment used in solid state NMR we utilize a frequency basis together with the spin basis to represent the Hamiltonian. In contrast to the standard Floquet theory we use a continuous frequency basis, which enables the description of an arbitrary time-modulated Hamiltonian. Nonetheless, we obtained a similar description, employing a continuous Fourier basis and Fourier transformations instead of the discrete Fourier basis and Fourier series as in the Floquet approach. Since we do not have a discrete Fourier basis, we do not obtain a representation of the Hamiltonian as a matrix of constant coefficients, but as a frequency dependent matrix function. However, the commutation relations are similar to those of the Floquet approach and allow for a similar treatment for example using van Vleck perturbation theory. \cite{VanVleck:1929fp,Primas:1961vi,Primas:1963tg}
%
%
\subsubsection{Generalization to multiple time-dependent modulations}
In principle, with the continuous frequency basis, there is no need to describe multiple time-dependent modulations using different frequency dimensions as is required in standard Floquet theory. All time-dependent modulations could be lumped into a single dimension. However, to understand resonance conditions between different processes, e.g., magic-angle spinning and radio-frequency irradiation of the spins, it is advantageous to separate this processes in different dimensions represented by multiple frequency bases. Let us generalize the approach presented in the previous section to $n$ time-dependent modulations of the Hamiltonian. As before we can describe the Hamiltonian in a frequency domain by using Fourier transformation, resulting in the expression 
\begin{align}
\HH(t) = \tau^n \int\limits_{-\infty}^{\infty} \int\limits_{-\infty}^{\infty}...\int\limits_{-\infty}^{\infty}  \widehat{\HH}(\Omega_1,\Omega_2,\cdots,\Omega_n) \notag\\ e^{i\Omega_1 t}  e^{i\Omega_2 t} \cdots e^{i\Omega_n t} \, \dd \Omega_1\, \dd \Omega_2 \cdots \dd \Omega_n
\label{eq:multimodal_Hamiltonian}
\end{align}
Now the underlying basis is the direct product basis of $n$ frequency bases and one spin basis  $\{\ket{\Omega_1,\Omega_2,\cdots,\Omega_n,\mu}\}$. This approach is similar to multimodal Floquet theory, but is not limited to periodic modulations of the Hamiltonian. 
From Eq. (\ref{eq:multimodal_Hamiltonian}) we directly obtain the resonance condition
\begin{align}
\Omega_1^{(0)} + \Omega_2^{(0)} + \cdots + \Omega_n^{(0)} = 0
\end{align}
Following the same procedure as in the single mode case we obtain a time-independent Hamiltonian of the form 
\begin{align}
\HH_{F} &= \int \int \cdots \int \widehat{D}_{1}(\Omega_1)\otimes \widehat{D}_{2}(\Omega_2) \otimes \cdots \otimes \widehat{D}_{n}(\Omega_n)\notag\\
&\quad \otimes\widehat{\HH}(\Omega_1,\Omega_2,\cdots,\Omega_n)\,\, \dd \Omega_1 \dd \Omega_2 \cdots \dd \Omega_n 
\notag\\
&\quad\oplus \widehat{\Omega}_1 \oplus \widehat{\Omega}_2\oplus \cdots \oplus \widehat{\Omega}_n
\label{Flo2}
\end{align}
where $\widehat{D}_j(\Omega)$ with $j \in \{1,2,\cdots,n\}$  is the corresponding frequency-translation operator and $\widehat{\Omega}_j$ the frequency operator.
It is important to emphasise that in practice most likely two frequency domains will be sufficient, a spatial modulation, for example due to sample rotation, and a modulation of the spin system due to rf-field irradiation. In this case the bimodal approach will be sufficient. However, for rf-field irradiation addressing different spin species, like proton, nitrogen or electron spins, a higher modal approach might offer advantages. In principle, one can also combine the frequency-domain approach in one dimension (e.g., rf irradiation) with the traditional Floquet approach in a different dimension (e.g., sample rotation) where the Fourier series is a perfect description of the time-dependent Hamiltonian (\textit{vide infra}).
%
%
\subsection{Van Vleck perturbation theory and effective Hamiltonian}
In the following, we apply van Vleck perturbation theory \cite{VanVleck:1929fp,Primas:1961vi,Primas:1963tg} to the Floquet Hamiltonian defined in Eq. (\ref{Flo}) with the goal of obtaining effective Hamiltonians of different orders. The detailed derivation can be found in the SI and follows the treatment in Ref.\cite{Ernst:2005ic} 
As usual we split the Hamiltonian into two parts
\begin{align}
\HH_F = \HH_F^{(0)} + \varepsilon\, \HH_F^{(1)}
\label{eq:perturbed}
\end{align}
and  apply the van Vleck transformation given by 
\begin{align}
e^{S}\,\HH_F\, e^{S^\dagger} = \sum_{m=0}^{\infty} \frac{[S,\HH_F]_m}{m!} = \sum_{n=1}^\infty \varepsilon^n \Lambda_F^{(n)}
\label{eq:perturbation_series}
\end{align}
where the nested commutator is defined as $[S,\HH_F^{(0)}]_m = [S,[S,\HH_F^{(0)}]_{m-1}]$ with $[S,\HH_F^{(0)}]_0 = \HH_F^{(0)}$.
We choose $S$, such that $[\Lambda_F,\HH_F^{(0)}] = 0$ and proceed by expanding $S$ as a series
\begin{align}
S = \sum_{n=1}^{\infty} \varepsilon^n S^{(n)}
\end{align}
and inserting it into Eq. (\ref{eq:perturbation_series}), which leads to 
\begin{align}
[S^{(n)},\HH_F^{(0)}] = \Lambda_F^{(n)} - \Phi^{(n)}_F
\label{eq:commutator_eq}
\end{align}
with
\begin{align}
\sum_{j=1}^{\infty} \varepsilon^j\Phi^{(j)}_F  = &\sum_{m=2}^{\infty} \frac{[\sum_{l=1}^{\infty} \varepsilon^l S^{(l)},\HH_F^{(0)}]_m}{m!} \notag\\
&+ \sum_{m=1}^{\infty} \frac{ [\sum_{l=1}^{\infty}\varepsilon^l S^{(l)},\varepsilon\, \HH_F^{(1)}]_m}{m!}.
\label{eq:Phi}
\end{align} 
Adopting the approach from Primas \cite{Primas:1961vi,Primas:1963tg} we obtain the formal solution of Eq. (\ref{eq:S}) as
\begin{align}
S^{(n)} &= \Gamma_F^{-1}(\Pi(\Phi^{(n)}_F) - \Phi^{(n)}_F) \label{eq:S}\\
\Lambda_F^{(n)} &=\Pi(\Phi^{(n)}_F) \label{eq:Lambda}
\end{align}
where $\Pi(X)$ is the projection operator
and $\Gamma_F^{-1}$ the inverse commutation operator defined in the SI.\\
Let us calculate the first and second-order effective Hamiltonians for the single-mode case:
\begin{align}
\HH^{(0)}_F &= \widehat{\Omega}\\ 
\HH^{(1)}_F & = \int \widehat{D}(\Omega)\otimes\widehat{\HH}(\Omega)\, \dd \Omega= \Phi^{(1)}_F
\end{align}
For the first-order effective Hamiltonian we evaluate Eq. (\ref{eq:Lambda}) leading to
\begin{align}
\bar{\HH}^{(1)} = \bra{\Omega'}\Lambda_F^{(1)}\ket{\Omega} = \widehat{\HH}(0) = \frac{1}{2\,\pi\,\tau}\int\limits_{0}^{\tau}  {\HH}(t)\, \dd t
\end{align}
For the second-order Hamiltonian we use Eqs. (\ref{eq:Phi}-\ref{eq:Lambda}) resulting in 
\begin{align}
\bar{\HH}^{(2)} = \frac{1}{2} \, PV
\int
\frac{[\widehat{\HH}(\Omega)
	,\widehat{\HH}(-\Omega)]}
{\Omega}
\dd\Omega
\label{eq:eff2}
\end{align}
We use the Cauchy principal value ($PV$) for the regularization of the integral, which avoids the integration over the singularity using limits. 
The main difference of Eq. (\ref{eq:eff2}) to Floquet theory is the integral over the frequency-domain Hamiltonian, instead of a sum of Fourier coefficients of the Hamiltonian. This disparity stems from the fact that the underlying Fourier space is continuous in contrast to the discrete space we use to describe periodic Hamiltonians.
\\
The derivation of the effective Hamiltonians for multiple frequency dimensions is quite similar and leads to the first and second-order effective Hamiltonian
\begin{align}
 \bar{\HH}^{(1)} &=  \int \widehat{\HH}(\Omega^{(0)}_1,-\Omega^{(0)}_1)d\Omega^{(0)}_1
\end{align}
and
\begin{align}
\bar{\HH}^{(2)} &= \frac{1}{2} PV
\int \dd\Omega_1 
\int \dd\Omega_1^{(0)}
\int \dd\Omega_2\notag\\
&\quad\frac{[\widehat{\HH}(\Omega_1,\Omega_2)
	,\widehat{\HH}(\Omega_1^{(0)}-\Omega_1,-\Omega_1^{(0)}-\Omega_2)]}
{\Omega_1 + \Omega_2}
\end{align}
where we used the resonance condition
\begin{align}
  \Omega^{(0)}_2=-\Omega^{(0)}_1  
\end{align}
In fact, we can retrieve the effective Hamiltonians obtained from Floquet theory, considering a periodic Hamiltonian. Periodicity with period $\tau_\textrm{m}$, causes quantization in the frequency domain with the frequency $\omega_\textrm{m} = 2\,\pi/\tau_\textrm{m}$. Therefore the frequency-domain Hamiltonian is only non-zero at the harmonics $n\,\omega_\textrm{m}$. Expressing this fact by a Dirac comb we obtain
\begin{align}
\HH(t) &= \int\limits_{-\infty}^\infty \widehat{\HH}(\Omega)\, e^{i \Omega t} \, \dd \Omega \notag\\
&= \int\limits_{-\infty}^\infty  \sum_{n=0}^{\infty}\delta(\Omega-n\,\omega_\textrm{m})\,\widehat{\HH}(\Omega)\, e^{i \Omega t} \, \dd \Omega \notag\\
&= \sum_{n=0}^{\infty} \widehat{\HH}(n\,\omega_\textrm{m})\, e^{i n \omega_\textrm{m} t} \notag\\
&= \sum_{n=0}^{\infty} \HH^{(n)}\, e^{i n \omega_\textrm{m} t}
\label{eq:trans_floquet}
\end{align}
where we can identify the Fourier coefficients of the Hamiltonian as $\widehat{\HH}(n\,\omega_\textrm{m})= \HH^{(n)}$.
This is simply the transition of the Fourier transformation to a Fourier series for a periodic function.  As a consequence  imposing a periodic boundary condition on the Hamiltonian leads to the results from Floquet theory.  
\\
As mentioned already above, it is possible to use a mixed approach, describing  periodic modulations with a discrete Fourier space and non-periodic modulations on a continuous Fourier space. We can use the same approach with the delta comb as in Eq. (\ref{eq:trans_floquet}) to arrive at
\begin{align}
\HH(t)&= \int\limits_{-\infty}^{\infty} \int\limits_{-\infty}^{\infty} \sum_{n} \delta(\Omega_1 - n \omega_\textrm{m})  \widehat{\HH}(\Omega_1,\Omega_2)\,e^{i\Omega_1 t}  e^{i\Omega_2 t} \, \dd \Omega_1\, \dd \Omega_2 \notag\\
&=  \int\limits_{-\infty}^{\infty} \sum_{n}  \widehat{\HH}(n \omega_\textrm{m},\Omega_2)\, e^{i n \omega_\textrm{m} t}  e^{i\Omega_2 t} \,\dd \Omega_2\notag\\
&=  \int\limits_{-\infty}^{\infty} \sum_{n}  \widehat{\HH}^{(n)}(\Omega_2) \, e^{i n \omega_\textrm{m} t}  e^{i\Omega_2 t} \,\dd \Omega_2
\label{eq:mixed}
\end{align}
This approach is useful to analyze general solid-state NMR experiments under MAS, where the sample spinning is described as a periodic modulation, but the rf-field irradiation can be arbitrary. Such mixed approaches can be used for any number of modes, where for example the periodic modulation due to the chemical shift offset is treated as a third mode.
The first and second-order effective Hamiltonians for the mixed approach with two frequencies as in Eq. (\ref{eq:mixed}) have the form
\begin{align}
\bar{\HH}^{(1)} &=  \sum_{n_0} \sum_{\Omega_0}  \widehat{\HH}^{(n_0)}(\Omega_0)
\end{align} 
and
\begin{align}
\bar{\HH}^{(2)} &=
\frac{1}{2}
\sum_{n} \sum_{n_0,\Omega_0}
PV \int \dd\Omega 
\quad\frac{[\widehat{\HH}^{(n)}(\Omega)
	,\widehat{\HH}^{(n_0-n)}(\Omega_0-\Omega)]}
{\Omega + n \omega_\textrm{m}}
\end{align}
with the resonance conditions defined by $\Omega_0+n_0\omega_\textrm{m}=0$.
%
Similar to standard Floquet theory, the propagation with an effective Hamiltonian is mediated by
\begin{align}
    \bar{\UU}(t) = \exp(- i\, \bar{\HH}^{(1)} \,t)
\end{align}
In contrast to standard Floquet Theory, the effective Hamiltonian is only valid for one specific duration, since the time dependence is implicitly in $\HH^{(1)}$. However, if the duration of the irradiation scheme is changed or the scheme is applied repeatedly, the effective Hamiltonian is weighted by a function, which depends on the duration or repetitions. In section \ref{sec:repeatedIrradiation} we derive this function and discuss its effects.
\subsection{Calculation of the frequency-domain interaction-frame trajectory}
As already mentioned, sample spinning in NMR leads to a periodic modulation of the Hamiltonian and can be described by a discrete Fourier series with the Fourier coefficients typically limited to the range -2 to 2. Radio-frequency irradiation is also often periodic in time, but typically an interaction-frame transformation is required to ensure the convergence of the effective Hamiltonian series. 
We split the Hamiltonian into two parts
\begin{align}
\HH(t) = \HH_\SSS(t) + \HH_\textrm{CS}(t) + \HH_\textrm{rf}(t) = \HH_\textrm{0}(t) + \HH_{1}(t)
\end{align}
where $\HH_\textrm{0}$ describes the spin-system Hamiltonian, $\HH_\textrm{rf}(t)$ is the rf-field Hamiltonian and $\HH_\textrm{CS}(t)$ the isotropic chemical-shift Hamiltonian. We can choose how to divide the total Hamiltonian into $\HH_0$ and $\HH_1$. Either we can set $\HH_1(t)=\HH_\textrm{rf}(t)$ and $\HH_0(t)=\HH_\SSS(t) + \HH_\textrm{CS}(t)$ resulting in an interaction-frame transformation by the rf irradiation only. In this case, the interaction-frame transformation is the same for all spins. Alternatively we can set $\HH_1(t)=\HH_\textrm{CS}(t)+\HH_\textrm{rf}(t)$ and $\HH_0(t)=\HH_\SSS(t)$ resulting in an interaction-frame transformation by the time-dependent effective field of each spin. In this case we have again two options. We can either include only the isotropic chemical shift into the interaction-frame transformation or we can include the isotropic and the anisotropic chemically shift. With this choice, the interaction-frame trajectory of each spin will be distinct if the chemical shifts are different. The selection of the most convenient interaction frame will depend on the system and problem at hand.
In the following we use an interaction frame generated by a general rf-field modulation including the chemical shift offset. In this case the unitary interaction-frame transformation of the Hamiltonian for each spin is a general complex rotation. Again, each spin will have its unique unitary transformation and, therefore, its unique frame, except if they are chemical equivalent. In the usual rotating frame defined by the Zeeman Hamiltonians, the Hamiltonian used for the interaction-frame transformation has the form
\begin{align}
\HH_{1}(t)
&= \omega_1(t) [\cos(\phi(t)) \II_x + \sin(\phi(t)) \II_y] + \omega_z(t) \II_z\notag\\
&= \vec\theta(t) \vec{\II}
\label{Hrf}
\end{align}
with
\begin{align}
&\vec{\theta}(t)
=
\begin{pmatrix}
\theta_x(t)\\
\theta_y(t)\\
\theta_z(t)\\
\end{pmatrix}
=
\begin{pmatrix}
\omega_1(t)\cos\left(\phi(t)\right)\\
\omega_1(t)\sin\left(\phi(t)\right)\\
\omega_z(t)
\end{pmatrix}\label{eq:rotation-vector}\\
&|\theta(t)| = \sqrt{\omega^2_1(t)+\omega^2_z(t)}
\end{align}
where $\omega_1(t)$ is the amplitude and $\phi(t)$ the phase of the rf-field and $\omega_z(t)$ the chemical shift offset. 
The direction of $\vec{\theta}(t)$ specifies the axis of rotation, its length $|\theta(t)|$ is the angular velocity at time $t$.
The interaction-frame transformation is given by
\begin{align}
\textrm{U}(t) = {\cal{T}}\exp\left(-i\,\int\limits_{0}^{t}\HH_1(t') \,\dd t' \right)
\label{general U}
\end{align}
where $\cal{T}$ is the Dyson time-ordering operator. \cite{Dyson:1949tf}
The evolution of the initial spin operators $\vec{\II} =
(\II_x,\II_y,\II_z)^T$ can be written with a single rotation matrix with elements $a_{\mu\nu}(t)$ as
\begin{align}
\II_\mu(t) = \UU^\dagger(t)\,\II_\mu\,\UU(t)
= \sum\limits_{ \nu}a_{\mu\nu}(t)\,\II_\nu
\end{align}
This also can be expressed in the frequency domain using Fourier
transform leading to
\begin{align}
&\II_\mu(t) =\Fou^{-1}\{\widehat{\II}_\mu(\Omega)\}= \sum\limits_{ \nu}a_{\mu\nu}(t)\,\II_\nu\\
&\widehat{\II}_\mu(\Omega)  = \Fou\{\II_\mu(t)\} =
\sum\limits_{ \nu}\widehat{a}_{\mu\nu}(\Omega)\,\II_\nu
\end{align}
where  $\widehat{a}_{\mu\nu}(\Omega) = \Fou\{a_{\mu\nu}(t)\}$.
In the following we present an efficient route to calculate $a_{\mu\nu}(t)$. Notice that the map $\II_\mu \mapsto \UU(t)\,\II_\mu\,\UU^\dagger(t)$ and $\II_\mu \mapsto(-\UU(t))\,\II_\mu\,(-\UU^\dagger(t))$ leads to the same rotation of $\mathbb{R}^3$, hence two elements of $SU(2)$ are mapped onto one element of $SO(3)$.
In more technical terms this means that there exists a 2:1 surjective homomorphism from $SU(2)$ to $SO(3)$.
However Eq. (\ref{general U}) yields to only one of the two $SU(2)$ elements, since it is impossible to get $-\UU(t)$ from a given $\UU(t)$ with the Hamiltonian of the form given in Eq. (\ref{Hrf}). Therefore, the operator defined in Eq. (\ref{general U}) is always a member of $SU(2)/Z_2$ which is isomorphic to $SO(3)$, i.e. $SU(2)/Z_2 \cong SO(3)$. This fact will be important for the back-calculation, when we infer the rf-irradiation from rotation matrices.A second important point is, that in general, $\HH_1(t)$ does not commute with itself at different times. For this reason we divide the irradiation into time intervals, small enough to assume that the Hamiltonian commutes with itself during the interval. However, the intervals do not necessary have to be of the same size and can be adjusted to fit the problem at hand. 
The propagator of the $j\,$th interval corresponds to $\UU_j$ and has a complex $2\times 2$ matrix representation given in the SI. Equipped with this matrix representation we calculate the coefficients of the rotation matrices $a^{(j)}_{\mu\nu}$ corresponding to $\UU_j$ as
\begin{align}
\UU_j^{\dagger}\,\II_\mu\,\UU_j = \sum_{\nu} a_{\mu\nu}^{(j)}\II_\nu \label{eq:single_rot}
\end{align}
Using $a_{\mu\nu}=2\, \text{Tr}(\II_\mu\, \UU\, \II_\nu\, \UU^{-1})$, leads to \cite{Cornwell:1984ws} 
\begin{align}
&a_{xx} = 
\left[\left(\theta^2_y+\theta_z^2\right)\cos \left(|\theta|\right)+\theta^2_x\right]/(4|\theta|)
\notag\\
& a_{xy} = \left[- \theta_x\theta_y
\left(\cos \left(|\theta|\right)-1\right)-\theta_z
|\theta| \sin \left(|\theta|\right)\right]/(4|\theta|)
\\
& a_{xz} = \left[
\theta_y
|\theta|
\sin \left(|\theta|\right)+ \theta_x
\theta_z\left(1- \cos
\left(|\theta|\right)\right)\right]/(4|\theta|)
\notag\\[0.75em]
& a_{yx} =\left[ \theta_z |\theta| \sin
\left(|\theta|\right)-\theta_x\theta_y \left(\cos
\left(|\theta|\right)-1\right)\right]/(4|\theta|)
\notag\\
& a_{yy} = \left[\left(\theta^2_x+\theta_z^2\right)\cos
\left(|\theta|\right) +\theta^2_y\right]/(4|\theta|)\label{eq:as}
\\
& a_{yz} = \left[ - \theta_y\, \theta_z \left(\cos
\left(|\theta|\right)-1\right)-\theta_x
|\theta| \sin \left(|\theta|\right)\right]/(4|\theta|)
\notag\\[0.75em]
& a_{zx} = \left[-\theta_y |\theta| \sin
\left(|\theta|\right)- \theta_z \theta_x \left(\cos
\left(|\theta|\right)-1\right)\right]/(4|\theta|)
\notag\\
& a_{zy} = \left[ \theta_x
|\theta| \sin
\left(|\theta|\right)-\theta_z \theta_y
\left(\cos \left(|\theta|\right)-1\right)\right]/(4|\theta|)
\\
& a_{zz} = \left[(\theta^2_x+\theta^2_y) \cos
\left(|\theta|\right)+\theta_z^2 \right]/(4|\theta|)
\notag
\end{align}
where we omitted the indices $j$ and $\theta$ on $a_{\mu\nu}$ for the sake of simplicity. Equivalent expressions can also be found for the alternate basis with $\mu,\nu\in \{+,-,z\}$.
The interaction-frame trajectory at time $t_j$ is given by
\begin{align}
{\bf{a}}(t_j)= {\bf{a}}_j \,{\bf{a}}_{j-1}\cdots{\bf{a}}_{1}
\label{spin-operator decomposition}
\end{align}
Note, that the rotation matrices ${\bf{a}}_j$ describe the rotation during a single time step while the matrix ${\bf{a}}(t_j)$
describes the total rotation up to the time point $t_j$.
The expressions for the elements $a_{\mu\nu}(t_j)$ in dependence on the parameter $\omega_1$,  $\omega_z$, and $\phi$ can be found in the SI. As mentioned before, the elements 
$\widehat{a}_{\mu\nu}{({\Omega})}$ are obtained by Fourier transformation of the elements $a_{\mu\nu}(t)$
\begin{align}
\widehat{a}_{\mu\nu}(\Omega) =\Fou\{a_{\mu\nu}(t)\}= \frac{1}{2\pi\tau} \int\limits_{0}^{\tau} a_{\mu\nu}(t) e^{i \Omega t} \, dt
\end{align}
The mathematical properties of $a_{\mu\nu}(t)$ and $\widehat{a}_{\mu\nu}(\Omega)$ can be summarized as follows
\begin{align}
a_{\mu\nu}(t) \in \mathbb{R} &\Leftrightarrow
\widehat{a}_{\mu\nu}(\Omega) = \widehat{a}^*_{\nu \mu}(-\Omega)\\
\sum_{\chi}a_{\mu\chi}(t)a_{\nu\chi}(t) = \delta_{\mu\nu} & \Leftrightarrow \sum\limits_{\chi} \widehat{a}_{\mu \chi}(\Omega)*\widehat{a}_{\nu \chi}(\Omega)=  \delta_{\mu\nu}\delta(\Omega)
\end{align}
%
%
%
%
\subsection{Properties of the interaction-frame trajectory}
\label{seq:porperties_trajectory}
In the following section the properties of the
time and frequency-domain interaction-frame trajectory are explored.
We will show that any finite irradiation can be expressed in term of
its periodic version, on one hand reducing the computational
effort, on the other hand isolating the effect of the finite duration.
Furthermore we will show that for any cyclic irradiation the calculation of the
frequency-domain interaction-frame trajectory can be reduced to the
calculation of a single segment, further lessening the computational efforts
significantly. Finally we will discuss the case of irradiation along a
single axis, which always enables the expression of the frequency-domain
interaction-frame trajectory in a closed form.
\subsubsection{General properties of the interaction-frame trajectory}
From the Plancherel theorem we obtain the general property \cite{Plancherel:1910bb}
\begin{align}\label{eq:planch}
  \frac{1}{2\pi\tau}\int\limits_{-\infty}^{\infty} |\widehat{a}_{\mu\nu}(\Omega)|^2\,d\Omega=
  \tau \int\limits_{0}^{\tau} |\widehat{a}_{\mu\nu}(t)|^2\,dt
\end{align}
In contrast to standard Floquet Theory, the duration of the
irradiation scheme is incorporated implicitly in the frequency-domain
interaction-frame trajectory.
However any finite $a_{\mu\nu}(t)$ defined on
$t\in [0,T]$ can be made periodic ($t \in \mathbb{R}$) using the modulo function
\begin{align}\label{eq:periodiced}
\widetilde{a}_{\mu\nu}(t) = a_{\mu\nu}(t\,\textrm{mod}\,T)
\end{align}
where we indicate the periodized function using a tilde. For a
periodic interaction-frame trajectory the frequency-domain interaction-frame trajectory is
related to the Fourier coefficients 
\begin{align}\label{eq:transform-series}
 &\widehat{a}_{\mu\nu}(\Omega) = \sum_{n=-\infty}^{\infty} a_{\mu\nu}^{(n)} \delta(\Omega-n\,{2\pi}/{T})
\end{align}
This relation is the bridge between standard Floquet theory, which
utilizes Fourier series and continuous Floquet theory, which 
uses the Fourier transform. This correspondence only exists for the
periodic case where standard Floquet theory is valid.
Vice versa, a periodic function $\widetilde{a}_{\mu\nu}(t)$ can be made
finite by multiplication with the appropriate rectangular window function
$\Pi(t)$
\begin{align}\label{eq:make-finite}
&a_{\mu\nu}(t) = \widetilde{a}_{\mu\nu}(t) \cdot \Pi\left(t/T-1/2\right)\\
&\Pi\left(t/T-1/2\right) =
\begin{cases} 
1 &  0\leq t \leq T \\
0 & \hspace{0.3em}\text{else}  
\end{cases}
\end{align}
Eq. (\ref{eq:make-finite}) and the convolution theorem lead to they
frequency-domain interaction-frame trajectory 
\begin{align}\label{eq:sinc_convolution}
    \widehat{a}_{\mu\nu}(\Omega) &= \widetilde{a}_{\mu\nu}(\Omega) * \mathcal{F}\left\{\Pi\left(t/T-1/2\right)\right\}\notag\\
                                &= \widetilde{a}_{\mu\nu}(\Omega) *\frac{\textrm{sin}(\Omega\,T/2)}{\pi \, \Omega \, T}e^{-i\frac{\Omega}{2}T}
\end{align}
Notice, that Eq. (\ref{eq:sinc_convolution}) can be efficiently
implemented numerically via discrete Fourier transformation (DFT),
since $\widetilde{a}_{\mu\nu}(\Omega)$ is periodic. 
Furthermore, the slope of $\widehat{a}_{\mu\nu}(\Omega)$ is dictated
by $T$, as $\widetilde{a}_{\mu\nu}(\Omega)$ has sharp frequency
components, see Eq. (\ref{eq:transform-series}). As a consequence
sharp edges of $\widehat{a}_{\mu\nu}(\Omega)$ cannot be realized by
any irradiation with finite duration. 
We can go a step further and decompose $\widetilde{a}_{\mu\nu}(t)$ into
a product and subsequently use the convolution theorem
\begin{align}
\widehat{a}_{\mu\nu}(\Omega)&=\mathcal{F}\left\{\Pi(t/T) \cdot \prod\limits_{i}^n\, \widetilde{a}^{(i)}_{\mu\nu}(t) \right\}
\notag
\\
&= \mathcal{F}\left\{\Pi(t/T)\right\}* \widetilde{a}^{(1)}_{\mu\nu}(\Omega)*\widetilde{a}^{(2)}_{\mu\nu}(\Omega) *\ldots* \widetilde{a}^{(n)}_{\mu\nu}(\Omega) \label{product_rule1}   
\end{align}
This trick will be useful to construct pulse schemes, since we can
decompose the desired $\widehat{a}_{\mu\nu}(\Omega)$ with convolutions
and calculate the necessary $a^{(j)}_{\mu\nu}$ for each
part. Finally we obtain the desired trajectory $a_{\mu\nu}(t)$ by
calculating the product $\prod^n_j a^{(j)}_{\mu\nu}$. The ability
to describe sequences of different length and the consequences of
finite recoupling sequences is one of the advantages of the new
method.
Independent of the basis of the spin operator, $a_{\mu\nu}(t)$ are the
elements of a rotation matrix, hence
\begin{align}
\sum_{\chi}a_{\mu\chi}(t)a_{\nu\chi}(t) = \delta_{\mu\nu} & \Leftrightarrow \sum\limits_{\chi} \widehat{a}_{\mu \chi}(\Omega)*\widehat{a}_{\nu \chi}(\Omega)=  \delta_{\mu\nu}\delta(\Omega)
\end{align}
The presented properties of the interaction-frame trajectory hitherto
always hold. However some symmetries of $a_{\mu\nu}(t)$ and
$\widehat{a}_{\mu\nu}(\Omega)$ depend on the basis of the spin
operator.
In the hermitian basis $\{\II_x,\II_y,\II_z\}$ the interaction-frame
trajectory is real and therefore the elements of the frequency-domain
interaction-frame trajectory are hermitian functions
\begin{align}
\widehat{a}^*_{\mu\nu}(t)&=\widehat{a}_{\mu\nu}(t)\\ 
\widehat{a}^*_{\mu\nu}(-\Omega)&=\widehat{a}_{\mu\nu}(\Omega) 
\end{align}
In contrast in the basis $\{\II_+,\II_-,\II_z\}$ complex conjugation
flips the sign of the indices, resulting in 
\begin{align}
a^*_{\pm \pm}(t) &=  a_{\mp\mp}(t)  \,\quad\quad\textrm{and}\quad a^*_{\pm\mp}(t) \,\,= a_{\mp\pm}(t)   \label{eq:conditions1}\\
  a^*_{\pm \pm}(\Omega) &=  a_{\mp \mp}(-\Omega) \quad\textrm{and}\quad a^*_{\pm \mp}(\Omega) = a_{\mp \pm}(-\Omega) \\
  a^*_{z\pm}(t) &=  a_{z\mp}(t) \quad\textrm{and}\quad\widehat{a}_{z+}(\Omega) = \widehat{a}^*_{z-}(-\Omega)\label{eq:conditionsend}
\end{align}
\subsubsection{Repetitive irradiation - nonzero effective field}\label{sec:repeatedIrradiation}
Many pulse schemes in NMR are repetitive, i.e., consist of repeating
segments. In general a repetitive irradiation does not generate a cyclic
interaction-frame trajectory but will lead to an effective field after
each segment. In this section we consider a pulse scheme of
duration $T$, consisting of $N$ repeating segments of duration $\tau$.
The interaction-frame trajectory of such an irradiation can be written
as
\begin{align}
    &a_{\mu\nu}(t) =\notag\\ 
    &\begin{dcases}
        a_{\mu\nu}(t) & 0 \leq t \leq \tau\\
        \sum_{\chi}a_{\mu\chi}(t-\tau)a_{\chi\nu}(\tau) & \tau \leq t \leq 2\tau\\
        \sum_{\chi}a_{\mu\chi}(t-2\tau)a_{\chi\nu}(2\tau) & 2\tau \leq t \leq 3\tau\\
        \hspace{2cm}\vdots & \quad\quad\vdots\\
        \sum_{\chi}a_{\mu\chi}(t-(N-1)\tau)a_{\chi\nu}((N-1)\tau) & (N-1)\tau \leq t \leq N\tau
    \end{dcases}
\end{align}
Notice that $a_{\mu\nu}(N\tau) = [\mathbf{a}^N(\tau)]_{\mu\nu}$ where
$\mathbf{a}(\tau)$ is the matrix with the elements
$a_{\mu\nu}(t)$.  The rotation matrix $\mathbf{a}(\tau)$ represents
the rotation of the spin due to the effective field. As shown in
detail in the SI, we obtain for a frequency-domain interaction-frame trajectory 
\begin{align}
  \widehat{a}_{\mu\nu}(\Omega) &= \frac{1}{2\pi N \tau} \sum_{n=1}^{N} \sum_{\chi}
    \int_{(n-1)\tau}^{n\tau} a_{\mu\chi}(t-(n-1)\tau) \notag\\ &\hspace{3cm}  \times a_{\chi\nu}((n-1)\tau)e^{-i\Omega t}dt\\
    &=
    \sum_\chi \bar{a}_{\mu\chi}(\Omega)\, g^{(N)}_{\chi \nu}(\Omega)
\end{align}
with
\begin{align}\label{eq:g}
  g^{(N)}_{\mu \nu}(\Omega) \coloneqq \frac{1}{N} \sum_{n=0}^{N-1} e^{-i n \Omega \tau} a_{\mu \nu}(n\,\tau)
\end{align}
and
\begin{align}
    \bar{a}_{\mu\nu}(\Omega) &= \frac{1}{2\pi\tau} \int\limits_{0}^{\tau} {a}_{\mu\nu}(t)e^{-i\Omega t} dt
     \\
     &= \Fou\{\widetilde{a}_{\mu\nu}(t) \Pi\left(t/\tau-1/2\right)\}
     \\
    &=\Fou\{\widetilde{a}_{\mu\nu}(t)\}*\mathcal{F}\left\{ \Pi(t/\tau-1/2) \right\}
\end{align}
As a result the $\widehat{a}_{\mu\nu}(\Omega)$ can be calculated
efficiently using DFT, since $a_{\mu\nu}(t)$ can be reduced to the periodized
trajectory of a single segment.
\subsubsection{Cyclic interaction-frame trajectory - zero effective field}
As shown previously, a cyclic interaction-frame trajectory $\mathbf{a}(t)$ can
be rewritten using modulo and a rectangular window function as
\begin{align}
{a}_{\mu\nu}(t) 
= a_{\mu\nu}(t\,\textrm{mod}\,\tau)\, \Pi\left(\frac{t-T/2}{T}\right) 
\label{eq:cyclic_trajectory}
\end{align}
The effective field for a cyclic interaction-frame trajectory after a
segment is always zero, i.e., the interaction frame ends at the same point where it started. As a consequence $\mathbf{a}(N\tau)=\mathbf{1}$, which leads
to 
\begin{align}
  g^{(N)}_{\mu \nu}(\Omega) =  \frac{1}{N} \sum_{n=0}^{N-1} e^{-i n \Omega \tau}
\end{align}
Notice, that in the case of a periodic interaction-frame trajectory,
i.e., an infinite repetition of the basic pulse scheme, the relation in
Eq. (\ref{eq:transform-series}) connects the
standard Floquet theory with continuous Floquet theory.
\subsubsection{Single irradiation axis}\label{sec:single_rf}
In many pulse schemes in NMR, rf irradiation is always along the same
axis. Therefore, the axis of rotation does not change during the pulse
scheme if we do not include the chemical-shift offset into the
interaction-frame calculation. In this case, the phase of the rf-field
Hamiltonian can be kept constant whereas the amplitude $\omega_1$ can
be positive or negative and allows rotations in both directions. A
special feature of the resulting rf-field Hamiltonian is that it
commutes with itself at different times and, therefore, an analytical
expression for the frequency trajectory $a_{\mu\nu}(\Omega)$ can be
found.
Without loss of generality, we consider the case where $\phi = 0$ and,
therefore, $\HH_\rf$ only generates rotation around the x-axis.  
It is convenient to transform in a tilted frame, where the z-axis is
aligned with the rotation axis of the rf-field.
The propagator resulting from $\HH_\rf$ together with the tilted-frame
transformation is
\begin{align}
\UU(t) = e^{-i\int\limits_{0}^{t}\omega_1(t)\dd t\, \II_z} e^{i \frac{\pi}{2} \II_y}
\end{align}
In most cases we only have to calculate the evolution of the spin
operator along the static magnetic field, which typically is along the
z-axis 
\begin{align}
\II_z(t) &= \UU^{\dagger}(t)\II_{z} \UU(t)\notag\\
&= \frac{1}{2}\left(e^{ i \int\limits_{0}^{t}\omega_1(t') dt'}\,\II^+ + e^{-i \int\limits_{0}^{t}\omega_1(t') dt'}\,\,\II^- \right)
\notag\\
&=\frac{1}{2} \left(a_{z+}(t)\,\II^+ + a_{z-}(t)\,\II^- \right)
\notag\\
&= \frac{\tau}{2} \int\limits_{-\infty}^{\infty} \left(a_{z+}(\Omega)\,\II^+ + \widehat{a}_{z-}(\Omega)\,\II^- \right)e^{i \Omega t} \dd\Omega
\label{eq:backcalc}
\end{align}
As a result, we get a relation between $\omega_1(t)$ and
$\widehat{a}_{z\pm}(\Omega)$
\begin{align}
&a_{z\pm}(t) =e^{\pm i \int\limits_{0}^{t}\omega_1(t') dt'} = \tau \int\limits_{-\infty}^{\infty} \widehat{a}_{z\pm}(\Omega) e^{i \Omega t}\,d\Omega 
\label{master11}\\
&\widehat{a}_{z\pm}(\Omega) = \frac{1}{2\pi \tau}\int\limits_{-\infty}^{\infty} e^{\pm i \int\limits_{0}^{t}\omega_1(t') dt'} e^{-i \Omega t}\,dt
                     \label{master2}
\end{align}
\\
We can solve Eq. (\ref{master11}) to obtain a solution for
$\omega_1(t)$  
\begin{align}
\omega_1(t) &= 
\mp i\, \frac{d}{dt} \ln\left(\Fou^{-1}\{\widehat{a}_{z\pm}(\Omega)\}\right) \notag \\
&= \pm  \frac{\mathcal{F}^{-1}\{\Omega \, \widehat{a}_{z\pm}(\Omega)\}}{\mathcal{F}^{-1}\{\widehat{a}_{z\pm}(\Omega)\}}\label{w1}
\end{align}
In addition, from the convolution theorem we obtain  
\begin{align}
\widehat{a}_{\mu\nu}^{(1)}(\Omega)*\widehat{a}_{\mu\nu}^{(2)}(\Omega) *\ldots* \widehat{a}_{\mu\nu}^{(n)}(\Omega) 
&= 
\mathcal{F}\biggr\{\prod\limits_{j=1}^n\, a_{\mu\nu}^{(j)}(t)\biggl\}\notag\\
=
\mathcal{F}\biggr\{\exp\biggr(&-i\,\int_{0}^{t}\sum\limits_{j=1}^{n}\omega_{1}^{(j)}(t) \dd t\biggl)
\biggl\}
\end{align}
This means that, if we decompose the desired shape
$\widehat{a}_{\mu\nu}(\Omega)$ as a convolution, we can just add the
corresponding $\omega_1(t)$ to obtain the desired rf-field profile.
%
%
\begin{figure*}
	\includegraphics[width=1\textwidth]{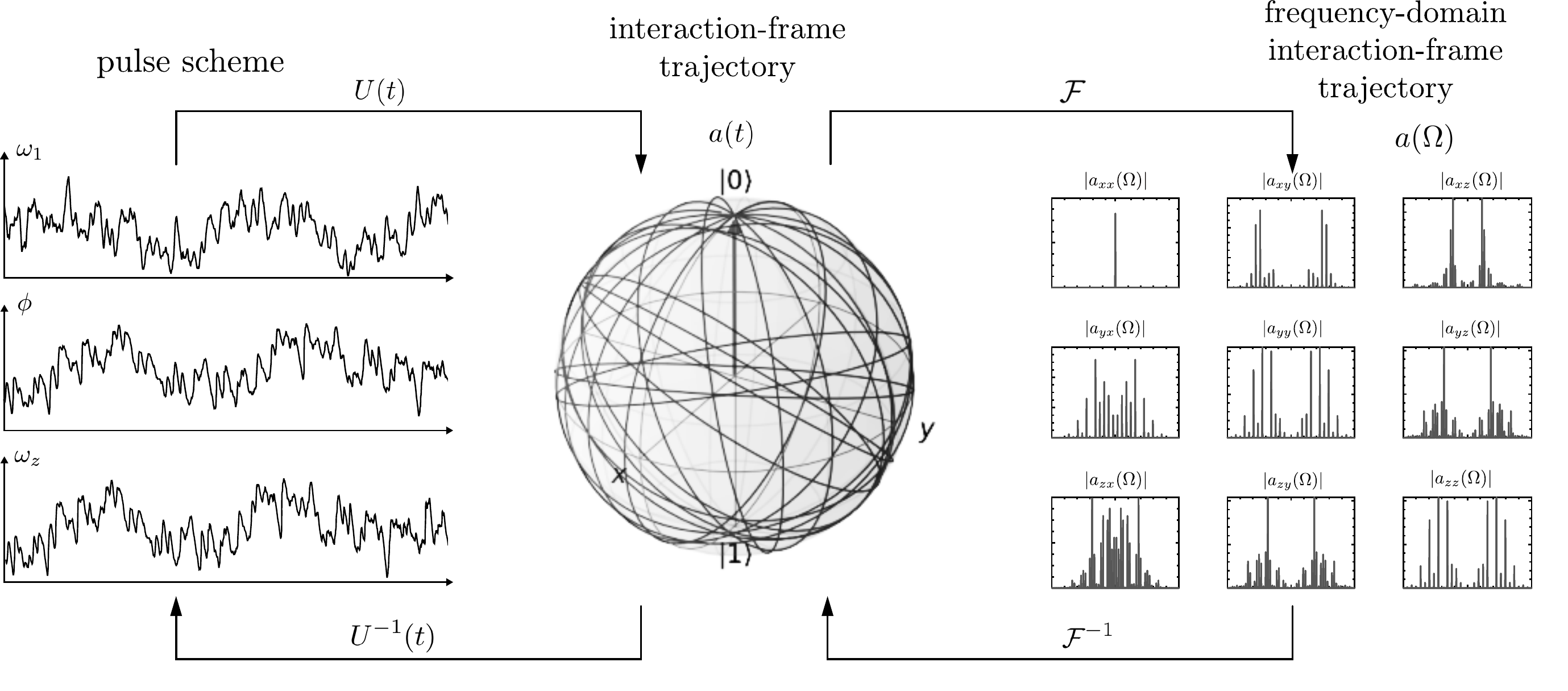}
	\caption{Schematic diagram of continuous Floquet theory for magnetic resonance. The information of the interaction-frame transformation $\UU(t)$ is encoded in the frequency-domain interaction-frame trajectory $\widehat{\mathbf{a}}(\Omega)$ which is calculated from the interaction-frame trajectory $\mathbf{a}(t)$. Subsequently $\widehat{\mathbf{a}}(\Omega)$ is combined with the effective Hamiltonian.}
\label{fig:backcalc}
\end{figure*}
\subsection{Calculation of the effective-field Hamiltonian from the frequency-domain interaction-frame trajectory}
In this section we reverse the previous procedure and calculate the rf-field Hamiltonian from $\widehat{a}_{\mu\nu}(\Omega)$.
As a first step we simply apply the inverse Fourier transform
\begin{align}
a_{\mu\nu}(t) = \tau \int\limits_{-\infty}^{\infty} \widehat{a}_{\mu\nu}(\Omega) e^{i \Omega t} \, d\Omega
\end{align}
In the next step we want to construct $\UU(t)$ from ${\bf{a}}(t)$. As mentioned previously the {U(t)} given in Eq. (\ref{general U}) are members of $SU(2)/Z_2$, which is isomorphic to $SO(3)$. Hence we can map each rotation matrix ${\bf{a}}_j$ uniquely to $\UU_j$. For the construction, we use quaternions, \cite{Cartan:1966tt,Blumich:1985vl} since they can represent complex as well as real rotations. 
More precisely, we reformulate the rotation matrices as unit quaternions and subsequently represent them with spin matrices in order to read out the pulse parameter.\\

A unit quaternion is given by
$
q_j = u + j\, v + k\, w + l \,z 
$
with
$j\,k\,l=-1$
and 
$u^2 + v^2 + w^2 + z^2 = 1$.
With the Euler-Rodriguez formula, \cite{Rodrigues:1840wl,Cartan:1966tt} we express its components with the elements of the corresponding rotation matrix
\begin{align}
u &= \frac{1}{2}\sqrt{1+tr({\mathbf{a}})} \\
v &= \frac{1}{4\,a}(a_{zy}-a_{yz})	\\ 
w &= \frac{1}{4\,a}(a_{xz}-a_{zx})  \\
z &= \frac{1}{4\,a}(a_{yx}-a_{xy})
\end{align}
Notice, that there are different options to calculate a unit quaternion from the corresponding rotation matrix. It is important for the numerical evaluation to choose the option, where the denominator is not close to zero, in order to increase the precision. We can always represent a unit quaternion as
\begin{align}
q_j = \cos\left(\frac{|\theta| \delta t}{2}\right) + \frac{1}{|\theta|}(j\, \theta_x + k\, \theta_y + l\,\theta_z ) \sin\left(\frac{|\theta| \delta t}{2}\right)
\end{align}
where $\vec\theta(t)$ is defined in Eq. (\ref{eq:rotation-vector}). The norm of $\vec\theta(t)$ is the angular velocity and the direction specifies the rotation axis.
Next, we rewrite the unit quaternions to assemble $\UU_j$ in its exponential form.
Therefore, we represent the quaternion units with the spin matrices
\begin{align}
& j = 2 i\, \II_x,\hspace{0.3cm} k = 2 i\, \II_y,\hspace{0.3cm}l = 2 i\, \II_z
\end{align}
which leads to the desired representation
\begin{align}
q_j &= \mathbb{I}\, \cos\left(\frac{|\theta| \delta t}{2}\right) - i \frac{\vec{\theta}}{|\theta|} 2\, \vec{\II}\, \sin\left(\frac{|\theta| \delta t}{2}\right)\notag\\
&=\exp(- i \, \vec{\theta}\,\, \vec{\II}\,\delta t) = \exp(-i\, \HH_\rf\, \delta t)
\end{align}
Together with the definition of $\vec\theta$ in Eq. (\ref{eq:rotation-vector}) we obtain
\begin{align}
u &= \frac{1}{2}\sqrt{1+\tr(a)} 
=\cos\left(\frac{|\theta|\delta t}{2}\right)\\
v &= \frac{1}{4\,u}(a_{zy}-a_{yz}) 
=\frac{\omega_1 }{|\theta|} \cos(\phi)\, \sin\left(\frac{|\theta|\delta t}{2}\right) \\
w &= \frac{1}{4\,u}(a_{xz}-a_{zx}) 
=\frac{\omega_1 }{|\theta|} \sin(\phi)\, \sin\left(\frac{|\theta|\delta t}{2}\right)  \\
z &= \frac{1}{4\,u}(a_{yx}-a_{xy})
=\frac{\omega_z }{|\theta|} \, \sin\left(\frac{|\theta|\delta t}{2}\right)
\end{align}
The solution of these equations for the pulse parameter $\omega_1,\,
\phi,\, \omega_z$ are
\begin{align}
|\theta|\delta t &= 2 \, \arccos\left(u\right) \\
\omega_1 &= \frac{|\theta|\, w}{\sin(\phi)\, \sin\left(\frac{|\theta|\delta t}{2}\right)} \\
\phi &= \arctan\left(\frac{w}{v}\right) \\
\omega_z &=  \frac{|\theta|\,z}{\sin\left(\frac{|\theta|\delta t}{2}\right)} 
\end{align}
which can be also expressed using $a_{\mu\nu}$ as
\begin{widetext}
\begin{align}
&|\theta|\delta t =\arccos\left(\frac{1}{2}\sqrt{1+\tr({\bf{a}})}\right)\\ \label{eq:backcalc0}
&\phi =\arctan\left(\frac{a_{xz}-a_{zx}}{a_{zy}-a_{yz}}\right) \\
&\omega_1 =  -\frac{\text{arcsec}\left(2\,(\tr({\bf{a}})+1)^{-\frac{1}{2}}\right) (a_{xz}-a_{zx}+a_{yz}-a_{zy})^2 \csc \left(2\, \text{arcsec}\left(2\,(\tr({\bf{a}})+1)^{-\frac{1}{2}}\right)\right)}{\delta t \left[\left(\tr({\bf{a}})+1\right)\left((a_{xz}-a_{zx})^2+(a_{yz}-a_{zy})^2\right)\right]^{\frac{1}{2}}}\\
&\omega_z =\frac{2\, (a_{yx}-a_{xy})\, \text{arcsec}\left(2\,(\tr({\bf{a}})+1)^{-\frac{1}{2}}\right)}{\delta t
	\left[\left(\tr(\bf{a})+1\right)\left(3-\tr({\bf{a}})\right)\right]^{\frac{1}{2}}}
\label{eq:backcalc1}
\end{align}
\end{widetext}
Eq. (\ref{eq:backcalc0} - \ref{eq:backcalc1}) express the phase, rf-field amplitude and offset as function of the interaction frame trajectory $a_{\mu\nu}(t)$. In order to obtain the complete sequence, these equations have to be evaluated at every time step of $a_{\mu\nu}(t_j)$.

Figure \ref{fig:backcalc} depicts the procedure to obtain the frequency-domain interaction-frame trajectory from the pulse parameter and vice versa. 
The mapping between the pulse scheme parameter frequency-domain interaction-frame trajectory is bijective, hence we always can find an unique result in both directions. However, only if the conditions given by Eqs. (\ref{eq:conditions1}-\ref{eq:conditionsend}) are fulfilled by the frequency-domain interaction-frame trajectory, a physical solution for the pulse parameter can be found.
%
%
%
\definecolor{darkgray}{rgb}{0.184314, 0.309804, 0.309804}
%
%
\section{The design of band-selective MIRROR experiments}\label{seq:Mirror}
In this section we apply continuous Floquet theory to design the rf irradiation for a tunable PDSD-based sequence called AM-MIRROR. \cite{Scholz:2008hr,Wittmann:2014go,Wittmann:2016bl}  The sequence can be designed for broadband or band-selective polarization transfer even at fast MAS frequencies by creating side-bands in the zero-quantum spectrum. Using the continuous frequency-space Floquet formalism, we show how to calculate the irradiation scheme to achieve tailored zero-quantum side-bands without any need of numerical optimization, which is in general not possible with the standard Floquet approach. 

\subsection{Theoretical preliminaries}\label{preliminaries}
To analyze the MIRROR experiment it is sufficient to consider a $\SSS_2\II$ spin system  with irradiation on the $\II$ spin. In the standard rotating frame such a Hamiltonian has the form
\begin{align}
\HH(t) = \HH_S(t) + \HH_\rf(t)
\end{align} 
with 
\begin{align}
\notag
\HH_S(t) =& \sum\limits_{n = -2}^2 \omega_{S_1S_2}^{(n)}e^{i n \omega_{\rr} t}\,(3 \, \SSS_{1z}\SSS_{2z}-\vec{\SSS}_1 \vec{\SSS}_2) 
\\
&+\sum\limits_{p=1}^2\sum\limits_{n = -2}^2 \omega_{I_1S_p}^{(n)}e^{i n \omega_{\rr} t}\,2 \, \II_{z}\SSS_{pz}
\label{Mirror-Hamiltonian} \notag
\\
&+\sum\limits_{p=1}^2 \omega_{S_p}^{(0)}\SSS_{pz} 
\end{align}
and 
\begin{align}
  &\HH_\rf(t)= \omega_1(t)\, \II_{x}
\end{align}
Here, consider only isotropic the chemical shifts of the $\SSS$ spins, but not of the $\II$ spins. Furthermore we assume an amplitude-modulated rf-irradiation in the x-direction ($\phi = 0, 180^{\circ}$).
For simplicity, we tilt the frame of reference to align the z-axis with rf-irradiation axis as in section \ref{sec:single_rf}. In addition, we transform into an interaction frame with the isotropic chemical shift of the S spins.
As usual we also transform into an interaction frame with the rf irradiation on the I spins leading to a total interaction-frame transformation defined by:
\begin{align}
\UU(t)=&
\underbrace{\textcolor{black}{\exp\biggl(-i\int\limits_{0}^t \omega_1(t')dt'\, \textcolor{black}{\II_z}\biggl)}}%
_{\textrm{rf-irradiation}}
\textcolor{black}{\underbrace{\vphantom{\int\limits_{0}^t} \textcolor{black}{\exp\biggl(-i \frac{\pi}{2}\, \II_y\biggl)}}%
	_{\textrm{tilted-frame}}}
\\\notag
&\underbrace{\vphantom{\int\limits_{0}^t} \textcolor{black}{\exp{\biggl(i\sum\limits_{p=1}^2 \omega_{Sp}^{(0)}t\, \SSS_{pz}\biggl)}}}%
_{\textrm{chemical shift}}
\end{align}
Since the S-spin chemical shift and the MAS spinning lead to a continuous rotation around a single axis,
the interaction-frame Hamiltonian $\widetilde{\HH}(t) \coloneqq\UU(t)\HH(t)\UU^\dagger(t)$ can be written as:
\begin{align}
\widetilde{\HH}(t) = T \sum\limits_{n = -2}^2 \sum\limits_{\ell=-1}^1\, \int\limits_{-\infty}^\infty\HH^{(n,\ell)}(\Omega)\, e^{i n\omega_{\rr} t} e^{i \ell\Delta\omega_{\iso} t}
e^{i \Omega t}\, \dd \Omega
\end{align}
from where we can read out the MIRROR resonance condition
\begin{align}
&\Omega_0 + n_0\omega_r + \ell_0\Delta\omega_{\textrm{iso}} = 0
\\
&(\Omega_0,n_0,\ell_0) \in \mathbb{R} \times \{\textrm{-}2,\textrm{-}1,...,2\} \times \{\textrm{-}1,0,1\}
\end{align}
The dominant second-order effective Hamiltonian is
\begin{widetext}
\begin{align}\label{eq: interaction-frame Hamiltonian}
\widetilde{\HH}_{SS \otimes IS} = \frac{1}{4} \sum\limits_{n, \nu = -2}^2 \, \int\limits_{-\infty}^\infty &\biggl( \frac{\omega_{S_1S_2}^{(\nu)}(\omega_{I_1S_2}^{(n-\nu)}-\omega_{I_1S_2}^{(n-\nu)})}{\Delta\omega_{\iso}-\nu\omega_{\rr}} + \frac{\omega_{S_1S_2}^{(n-\nu)}(\omega_{I_1S_2}^{(\nu)}-\omega_{I_1S_2}^{(\nu)})}{\Delta\omega_{\iso}+\nu\omega_{\rr}} \biggr)
\,(\widehat{a}_{z\pm}(\Omega)\,\II^{\pm}+\widehat{a}_{z\mp}(\Omega)\,\II^{\mp}) \,\SSS_1^{\pm} \SSS_2^{\mp} \, d\Omega
\end{align}
\end{widetext}
where 
\begin{align}
\widehat{a}_{z\pm}(\Omega) &\coloneqq \widehat{a}_{zx}(\Omega) \pm i\,\widehat{a}_{zy}(\Omega)
\end{align}
In the following we are going to use Eq. \eqref{w1}
which enables the calculation of the rf-field amplitude.
%
%
%
\subsection{Targeting single $\Delta\omega_{\textrm{iso}}$}
The first example is the recoupling of a single chemical-shift difference $\Delta\omega_{\mathrm{iso}}$ given by
\begin{align}
\widehat{a}_{z+}(\Omega) \stackrel{!}{=} \delta(\Omega-\Delta\omega_{\textrm{iso}})
\hspace{2em} 
\end{align}
Inserting it in Eq. (\ref{master11}) leads to
\begin{align}
&e^{i \int\limits_{0}^{t}\omega_1(t') dt'}=\int\limits_{-\infty}^{\infty}\delta(\Omega-\Delta\omega_{\textrm{iso}}) e^{i \Omega t}\,d\Omega=e^{i \Delta\omega_{\textrm{iso}} t}
\end{align}
hence
\begin{align}
\omega_1(t) = \Delta\omega_{\textrm{iso}}
\end{align}
Alternatively we can use Eq. (\ref{w1})
\begin{align}
\omega_1(t) 
&=\frac{\mathcal{F}^{-1}\{\Omega \, \delta(\Omega-\Delta\omega_{\textrm{iso}})\}}{\mathcal{F}^{-1}\{\delta(\Omega-\Delta\omega_{\textrm{iso}})\}}
= \frac{\Delta\omega_{\textrm{iso}} e^{-i \Delta\omega_{\textrm{iso}} t}}{e^{-i \Delta\omega_{\textrm{iso}} t}}
= \Delta\omega_{\textrm{iso}}
\end{align}
This solution is not physical, since it would require irradiation of a infinite duration. 
As we will show below, the optimal rf-field amplitude for a given duration $T$ is in our case
\begin{align}
\omega_1(t)=
\begin{cases} 
\Delta\omega_{\textrm{iso}} &  \,0 \leq t \leq T \\
0 & \hspace{1.5em}else  
\end{cases}
\end{align}
Using Eq. (\ref{master2}), we obtain
\begin{align}
\widehat{a}_{z+}(\Omega) &= \frac{i}{2\pi T}\,\frac{
	1-e^{i \text{$T $} (\Omega -\text{$\Delta\omega_{\textrm{iso}}
			$})}}{
	\Delta\omega_{\textrm{iso}}-\Omega}
\end{align}
Note that this is a sinc function where the time origin was shifted to the beginning of the time period and not at the center.
The singularity at $\Omega =\Delta\omega_{\textrm{iso}}$ is removable, which can be seen from its expansion
\begin{align}
\widehat{a}_{z+}(\Omega) &= \frac{i}{2\pi}\, \sum\limits_{m=1}^\infty \frac{i^m}{m!}[T(\Delta\omega_{\textrm{iso}} -\Omega)]^{m-1}
\label{expansion}
\end{align}
We can easily calculate the global maximum of $|\widehat{a}_{z+}(\Omega)|$ with Eq. (\ref{expansion}), which is located at  $\Omega =\Delta\omega_{\textrm{iso}}$ 
\begin{align}
|a_{z+}(\Delta\omega_{\textrm{iso}})| &= \frac{1}{2\pi}
\end{align}
Notice, that the global maximum is independent of the duration of the pulse scheme, because of the chosen normalization.
\subsection{Targeting multiple isolated $\Delta\omega_{\textrm{iso}}$}
The simplest example of targeting multiple frequencies is to target 2 frequencies
\begin{align}
\widehat{a}_{z+}(\Omega) \stackrel{!}{=} \frac{1}{2}\left(\delta(\Omega-\Delta\omega_{\textrm{iso}}^{(1)}) + \delta(\Omega-\Delta\omega_{\textrm{iso}}^{(2)})\right)
\end{align}
Inserting it into Eq. (\ref{w1}) leads to 
\begin{align}
\omega_1(t) = \Delta\omega_{\textrm{iso}}^{(1)}+\Delta\omega_{\textrm{iso}}^{(2)}
\end{align}
This irradiation is not possible for 2 reasons. First, we would again require irradiation of infinite duration as before. The second problem is that two waves with the same phase just add up in amplitude, i.e. $\Delta\omega_{\textrm{iso}}^{(1)}+\Delta\omega_{\textrm{iso}}^{(2)} = \Delta\omega_{\textrm{iso}}^{(3)}$. As a result, we would just target one frequency, namely $\Delta\omega_{\textrm{iso}}^{(3)}$. Notice, that this is the only existing solution for the given $\widehat{a}_{z+}(\Omega)$. Therefore we have no choice than to split the irradiation into parts that target the two frequencies separately and successively in time
\begin{align}
\omega_1(t)=
\begin{cases} 
\Delta\omega_{\textrm{iso}}^{(1)} &  \,0 \leq t \leq \tau_1 \\
\Delta\omega_{\textrm{iso}}^{(2)} & \tau_1 < t \leq T  
\end{cases}
  \label{eq:step-irradiation}
\end{align}
We again can use Eq . (\ref{master2}) and obtain
\begin{align}
\widehat{a}_{z+}(\Omega) & =\widehat{a}^{(1)}_{z+}(\Omega)+\widehat{a}^{(2)}_{z+}(\Omega) \notag\\
&=\frac{i}{2\pi T}\frac{
	\left(1-e^{-i \text{$\tau_1 $} (\Omega -\text{$\Delta\omega_{\textrm{iso}}^{(1)}
			$})}\right)}{(
	\text{$\Delta\omega_{\textrm{iso}}^{(1)} $}-\Omega)} \notag
\\
&\quad +\frac{i}{2\pi T}\frac{
	\left(e^{-i \text{$\tau_1 $} (\Omega -\text{$\Delta\omega_{\textrm{iso}}^{(2)} $})}-e^{-i
		\text{$T $}
		(\Omega -\text{$\Delta\omega_{\textrm{iso}}^{(2)} $})}\right) e^{i (\text{$\tau_1 $} \text{$\Delta\omega_{\textrm{iso}}^{(1)}
			$}-\text{$T $} \text{$\Delta\omega_{\textrm{iso}}^{(2)} $})}}{ (\text{$\Delta\omega_{\textrm{iso}}^{(2)} $}-\Omega
  )}
  \label{eq:twolines}
\end{align}
Notice, that only the phase factor of the second term $a^{(2)}_+(\Omega)$ depends on the first part of the irradiation, but not its absolute value.
Similar to the case of a single chemical shift we find
\begin{align}
|\widehat{a}^{(1)}_{z+}(\Delta\omega^{(1)}_{\textrm{iso}})| &= \frac{\tau_1}{2\pi T}\\
|\widehat{a}^{(2)}_{z+}(\Delta\omega^{(2)}_{\textrm{iso}})| &= \frac{T-\tau_1}{2\pi T}
\end{align}
which approximates well the height of the two highest maxima of $|\widehat{a}_{z+}(\Omega)|$, becoming more accurate the bigger $|\Delta\omega^{(2)}_{\textrm{iso}}-\Delta\omega^{(1)}_{\textrm{iso}}|$, because the overlap between $a^{(1)}(\Omega)$ and $a^{(2)}(\Omega)$ decreases.
We can generalize this findings to $N$ chemical shifts as
\begin{align}
\omega_1(t)=
\begin{dcases}
\begin{aligned}
&\Delta\omega^{(1)}_{\textrm{iso}} \hspace{2.4em}  0 \leq t \leq \tau_1\\
&\Delta\omega^{(2)}_{\textrm{iso}} \hspace{2em}  \tau_1 < t \leq \tau_2\\[-1ex]
&\hspace{1em}\vdotswithin{\Omega} \hspace{5em}\vdots \\[-0.5ex]
&\Delta\omega^{(n)}_{\textrm{iso}}     \hspace{0.8em} \tau_{n-1} < t \leq \tau_n
\end{aligned}
\end{dcases}
\label{general_rf}
\end{align}
leading to
\begin{align}
\widehat{a}_{z+}(\Omega)  &=\frac{i}{2\pi T} \sum\limits_{n=0}^N \frac{\left(e^{-i
		\text{$\tau_{n-1} $} (\Omega
		-\text{$\omega_n$})} - e^{-i \tau_n (\Omega -\text{$\omega_n $})}\right)}{ (\omega_n-\Omega)}
\notag\\
& \quad e^{i (\tau_1 \omega_1 + \tau_2 \omega_2 + \cdots + \tau_{n-2} \omega_{n-2} + \tau_{n-1} (\omega_{n-1} -\omega_n))}\label{generalstep}
\end{align}
where we set $\tau_N = T$ and  $\tau_0 = 0 = \omega_0$.
Like before, the maxima of the absolute value of each term is given by  
\begin{align}
|\widehat{a}^{(n)}_{z+}(\Delta\omega^{(n)}_{\textrm{iso}})| &= \frac{\tau_{n+1}-\tau_n}{2\pi T}
\end{align}
\\
It might be possible to construct an arbitrary function from general stepwise function by taking the limits carefully.
The same limit can be applied directly on Eq. (\ref{generalstep}), if the integral in Eq. (\ref{master11}) is dominant integrable \cite{Bartle:2014th} and hence we can change the order of the limit and the intergral operation.
\newpage
\subsection{Targeting a range of $\Delta\omega_{\textrm{iso}}$}\label{Targeting a range}
To obtain $\omega_1(t)$ we set $\widehat{a}_{z+}(\Omega)$ to be a rectangular function with width $b$, normalized according to Eq. (\ref{eq:planch}), as
\begin{align}
\widehat{a}_{z+}(\Omega) \stackrel{!}{=}  \sqrt{\frac{2\pi}{b}}\,\Pi(\Omega/b)
\end{align}
From Eq. (\ref{w1}) follows
\begin{align}
\omega_1(t)
= \frac{\mathcal{F}^{-1}\{\Omega\, \,\Pi(\Omega/b)\}}{\mathcal{F}^{-1}\{\,\Pi(\Omega/b)\}}
= i \left(\frac{b}{2} \cot\left(\frac{b\,t}{2}\right)-\frac{1}{t} \right)
\end{align}
Notice that $\omega_1(t)$ is independent of a constant factor of $\widehat{a}_{z+}(\Omega)$. 
As before we have the problem of infinite duration in order to obtain the desired shape exactly.\\
In contrast to the previous theoretical rf-field amplitudes, we get a time-dependent function, which has poles and is purely imaginary.
Motivated by Eq. (\ref{general_rf}) and Eq. (\ref{generalstep}), we use a linear function
\begin{align}
\omega_1(t)= \frac{(\Delta\omega_{\textrm{iso}}^{(2)}-\Delta\omega_{\textrm{iso}}^{(1)})}{\vphantom{T^5}T}\,t + \Delta\omega_{\textrm{iso}}^{(1)}\label{eq:linearfunction}
\end{align}
Inserting it in Eq. (\ref{master2}), yields
\begin{align}
\widehat{a}_{z+}(\Omega)&=\sqrt{\frac{\pi}{2T}}\, e^{i \frac{3\pi}{4}} e^{i\frac{T
		(\Delta\omega^{(1)}_{\textrm{iso}}-\Omega)^2}{2 (\Delta\omega^{(1)}_{\textrm{iso}}-\Delta\omega^{(2)}_{\textrm{iso}})}}
\notag\\
\times&\frac{\text{erf}\left(\frac{\left(\frac{1}{2}+\frac{i}{2}\right) \sqrt{T}
		(\Delta\omega^{(2)}_{\textrm{iso}}-\Omega)}{\sqrt{\Delta\omega^{(1)}_{\textrm{iso}}-\Delta\omega^{(2)}_{\textrm{iso}}}}\right)-\text{erf}\left(\frac{\left(\frac{1}{2}+\frac{i}{2}\right)
		\sqrt{T}
		(\Delta\omega^{(1)}_{\textrm{iso}}-\Omega)}{\sqrt{\Delta\omega^{(1)}_{\textrm{iso}}-\Delta\omega^{(2)}_{\textrm{iso}}}}\right)}{\sqrt{\Delta\omega^{(1)}_{\textrm{iso}}-\Delta\omega^{(2)}_{\textrm{iso}}}}
\label{eq:box}
\end{align}
Taking a closer look on Eq. (\ref{eq:linearfunction}) and Eq. (\ref{eq:box}), we notice that Eq. (\ref{eq:linearfunction}) consists of 2 parts. The first term causes a rectangular like shape, the second term shifts this shape to the desired position.
An intuitive picture offers Eq. (\ref{master11}) when we consider an interaction-frame rotation with an uniformly changing frequency $a_{z+}(t) = e^{i(m\,t + c)t}$. The frequencies $\omega(t) = m\,t + c$ are visible in its Fourier transform. Sweeping uniformly through a range of nutation frequencies slowly once, will drive more resonances than to sweep the same range several times, even if the overall time is the same. This is because a repeated sweep, also includes other, much slower, frequency components, due to the repetition.

\begin{figure}[h]
	\includegraphics[width= 0.5\textwidth]{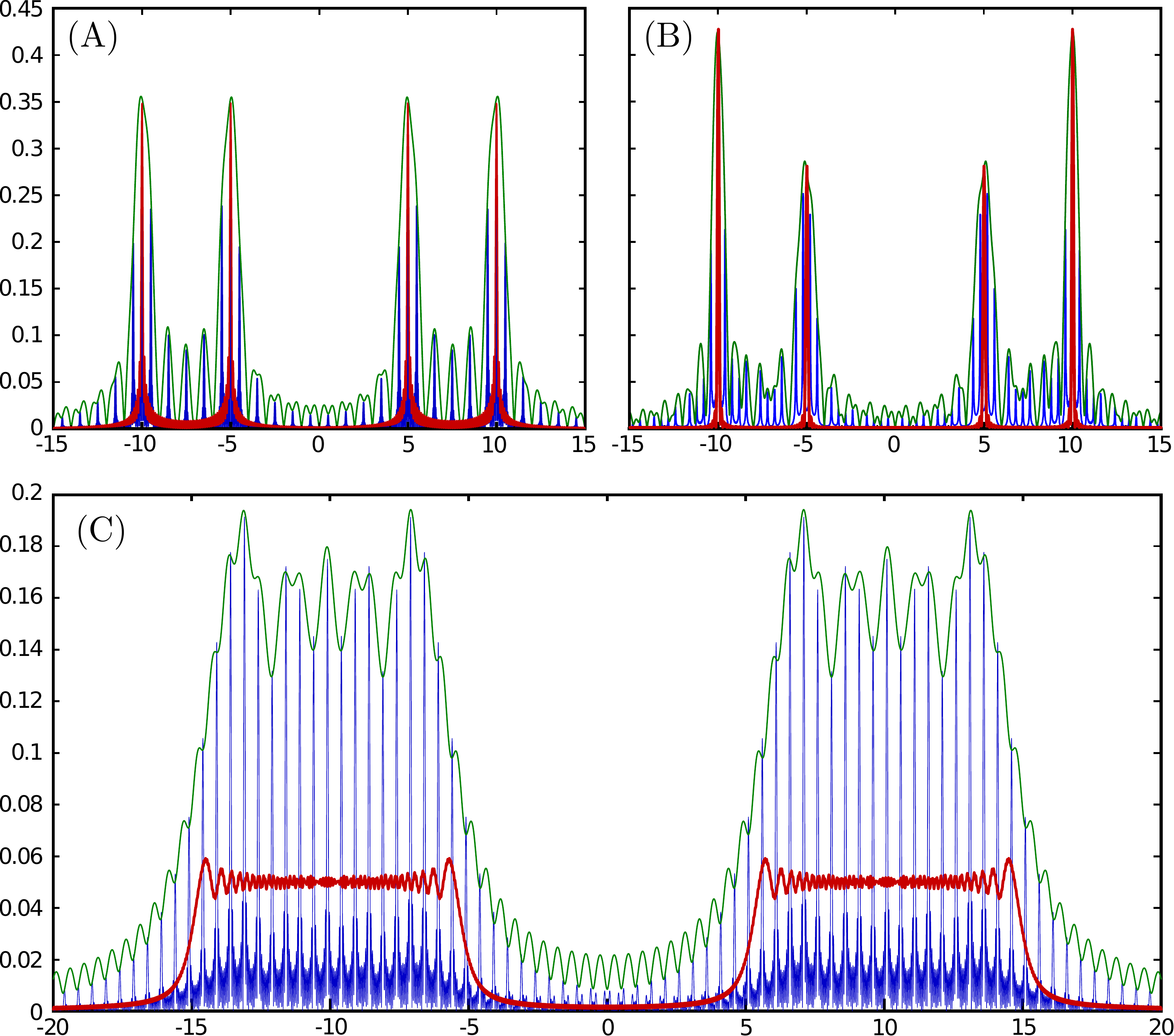}
	\caption{$\frac{1}{4\pi}\bigl( |\widehat{a}_{z+}(\Omega)|^2 + |\widehat{a}_{z-}(\Omega)|^2\bigr)$ for different irradiation schemes. (A) Step-wise irradiation (Eqs. (\ref{eq:step-irradiation}) and (\ref{eq:twolines})) with $\Delta \omega^{(1)}_{\iso} =5$ and  $\Delta \omega^{(2)}_{\iso} = 10$ with $T=2\tau_1=20\,$ms.  (B) Same as (A) but with $\tau_1 = 3/4\, T$. (C) Ramped irradiation described by Eq. (\ref{eq:linearfunction}) and Eq. (\ref{eq:box}), ranging from $7.5\,$kHz $\leq  \omega_1 \leq  12.5\,$kHz.  
For the red curves, the duration has been set to $T=20$, repeating the shape once.
For the blue curves the duration has been set to $T/10$ and repeated $10$ times. For the green curves the duration has been $T/10$ and repeated once, but subsequently scaled up $10$ times.    The integral of the red and blue curve is equal, since the duration of the irradiation is the same. As a consequence of the repetition, the blue curve is discretized by the modulation frequency $\nu_m = 0.5\,\textrm{kHz}$.}
\label{fig:repetition}
\end{figure}

Figure \ref{fig:repetition} shows $\frac{1}{4\pi}\bigl(|\widehat{a}_{z+}(\Omega)|^2 + |\widehat{a}_{z-}(\Omega)|^2\bigr)$ for different pulse schemes and different number of repetitions of the pulse scheme. For Figs.  (\ref{fig:repetition}A) and (\ref{fig:repetition}B) a step-wise pulse schemes (Eq. (\ref{eq:step-irradiation}) and Eq. (\ref{eq:twolines})) was used with $\tau_1=T/2$ for (A) and $\tau_1=3/4T$ for (B). For Fig. (\ref{fig:repetition}C) a ramped irradiation described with Eqs. (\ref{eq:linearfunction}) and  (\ref{eq:box}) was used. For the red curves the pulse scheme was applied only once during $T=20\,$ms leading to 4 individual peaks for (A) and (B) and a box-shaped curve for (C). For the blue curves the irradiation scheme has been applied 10 times during the same duration of $T=20\,$ms, introducing the modulation frequency $\tau_m = 2\,$ms. As a consequence the blue curves is discretized due to the modulation frequency $\nu_m=1/\tau_m = 0.5\,$kHz. The green curves is obtained by applying the irradiation scheme  once for $T=2\,$ms and scaling the curves by a factor of 10, which leads to the envelope of the blue curves.
Notice that all the curves shown were obtained by analytical expressions, i.e using Eq. (\ref{eq:twolines}) for  (A) and (B) and Eq. (\ref{eq:step-irradiation}) for  (C). To account for the effect of the repetition Eq. (\ref{eq:g}) was used, and therefore all plots are obtained by FFT of a single element.
\subsection{Comparison to numerical simulation}
\begin{figure}[h]
\includegraphics[width=0.4\textwidth]{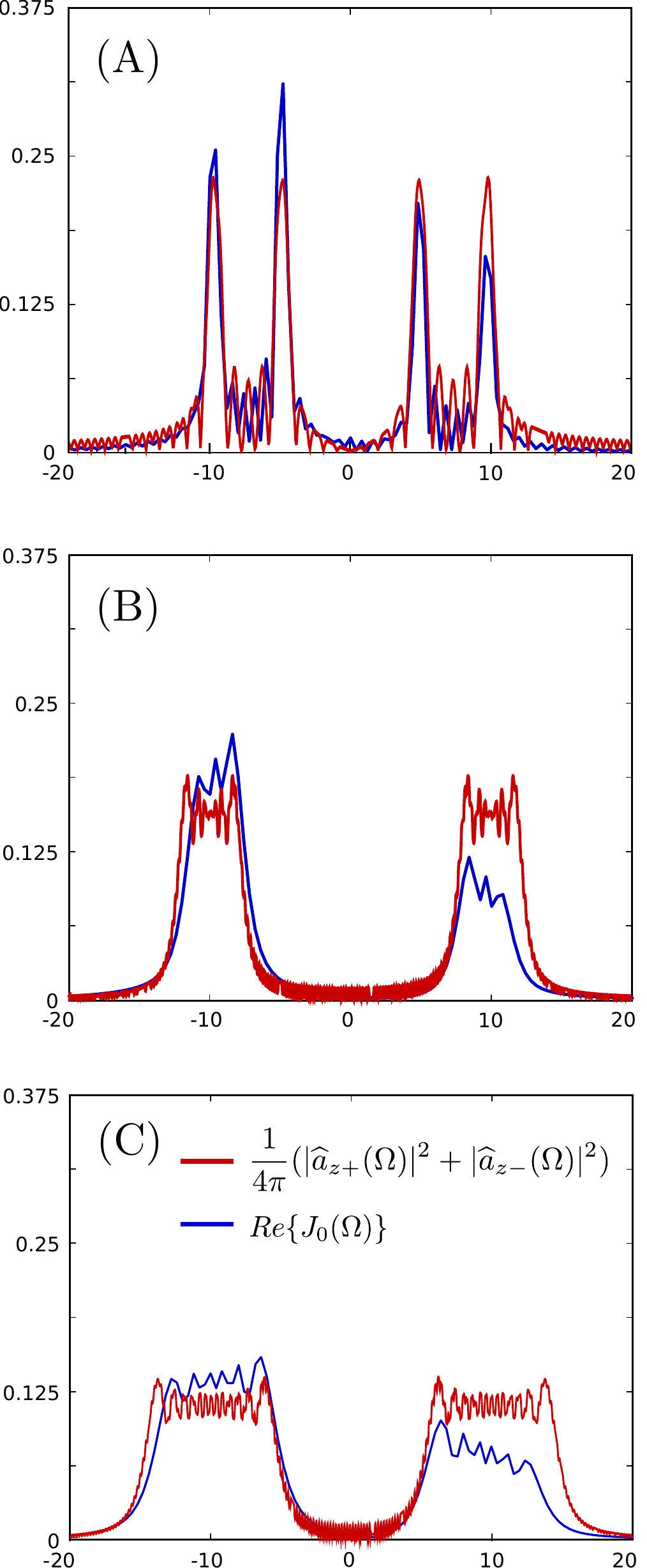}
\caption{Comparison between numerical simulation of the zero quantum line and the interaction frame trajectory for different irradiation schemes. (A) Step irradiation as given in Eq. (\ref{eq:step-irradiation}) with $\Delta\omega_{\text{iso}}^{(1)}= 5\,$kHz, $\Delta\omega_{\text{iso}}^{(2)}$ and $\tau_1=T/2=10$\,ms. (B) Ramp irradiation with $7.5\,$kHz $\leq \omega_1 \leq  12.5\,$kHz. (C) Ramp irradiation with $7.5\,$kHz $\leq \omega_1 \leq  12.5\,$kHz.}
\label{fig:comparionZQ}
\end{figure}

An alternate description of such PDSD-type  experiments which does not rely on effective Hamiltonians, was provided by  Veshtort \cite{Veshtort:2011iv}, where the rate constants $k_\mathrm{D}$ under MAS can be calculated as
\begin{align}
  \label{eq:PDSD-rate}
  k_\mathrm{D}=\sum^{2}_{n=-2} |\omega^{(n)}_{S_1S_2}|^2\, Re\{J_0(\Delta\omega_{\text{iso}}-n\omega_r)\}
\end{align}
Here $\omega^{(n)}_{S_1S_2}$ is defined as in Eq. (\ref{Mirror-Hamiltonian}) and  $J_0(\omega)$ is the ZQ line \cite{Veshtort:2011iv}. Notice, that $k_\mathrm{D}$ is proportional to $J_0(n\omega_\mathrm{r})$, which can be manipulated by the irradiation and is dependent on the MAS frequency, similarly to $\widehat{a}_{z\mp}(\Omega)$.
However the zero-quantum line and  $\widehat{a}_{z\mp}(\Omega)$ are quantities that stem from fundamentally different theoretical descriptions. Nevertheless both quantities ultimately describe the evolution of the ZQ operator and dictate the ZQ transfer selectivity and efficiency. 
Figure \ref{fig:comparionZQ} shows a comparison of the ZQ line and $1/(4\pi) (|\widehat{a}_{z+}(\Omega)|^2+|\widehat{a}_{z-}(\Omega)|^2) $ for different irradiation schemes. The plots of the ZQ line are the results of numerical time-slicing simulation of a CH$_2$ spin system using the C++ library GAMMA \cite{Smith:1994te}. The mixing time was set to $20\,$ms and the MAS frequency to $50\,$kHz. The  $\widehat{a}_{z\pm}(\Omega)$ were calculated from Eq. (\ref{eq:twolines}) and (\ref{eq:box}) using the same mixing time as in the numerical simulation. As mentioned before the two quantities stem from different descriptions, but the agreement, especially of the width and position, is very good, underlying their physical similarities.
\newpage
\section{Conclusion}
We presented a generalization of operator-based Floquet theory to non-periodic Hamiltonians.
Instead of a discrete frequency space, which is sufficient for the description of periodic Hamiltonians as in the standard Floquet approach, we utilized a continuous frequency space allowing non-periodic Hamiltonians with finite duration.
In contrast to periodic modulations, modulation with finite duration led to resonance conditions that are not infinitely sharp, enabling the description of non-resonant behaviors within the same framework, which is very cumbersome in standard Floquet theory.
Since the framework is not restricted to periodic modulations, it can be use in combination with any interaction-frame transformation. This might be useful to analyze the effect of specific terms in the Hamiltonian by choosing an appropriate interaction frame, or simplifying the description by incorporating all single-spin modulations in the interaction frame.
Despite the finite duration of the modulations, we can still utilize FFT for computationally efficient calculations and express the effect of the repetition of a pulse scheme by a simple quantity. 
We saw that the pulse parameter and the interaction-frame trajectory, which encodes the modulation of the spin system, can be mapped to each other bijectively, enabling reverse engineering pulse schemes from the effective Hamiltonian to a certain extent.
Finally we applied the formalism to a PDSD-based sequence called AM-MIRROR where we could tailor desired zero-quantum side-bands using the mentioned bijective mapping, without any need of numerical optimization. 
This framework was developed to describe and design solid state NMR experiments under MAS, but is not restricted to it. It can be used to describe experiments of other magnetic resonance technologies such as DNP and EPR.

\section*{ACKNOWLEDGMENTS}
This research has been supported by the ETH Zürich and the Schweizerischer Nationalfonds zur Förderung der Wissenschaftlichen Forschung (grant no. 200020\_188988).

\section*{AUTHORS DECLARATION}

\subsection*{Conflict of interest}
The authors have no conflicts of interest to disclose.


\section*{DATA AVAILABILITY}
The simulated data will be uploaded to a public repository after acceptance of the paper.

\nocite{*}
\bibliography{continuous_floquet.bib}

\begin{widetext}
\section{Supplementary Information}
\section{Derivation of Van Vleck perturbation theory on a
  continuous frequency space} \label{sec:van-vleck}
In the following we apply Van Vleck perturbation theory on the single-mode Floquet Hamiltonian.
As usual we split the Hamiltonian into two parts
\begin{align}
\HH_F = \HH_F^{(0)} + \varepsilon\, \HH_F^{(1)}
\label{eq:perturbed}
\end{align}
Next we apply a unitary transformation called van Vleck transformation on Eq. (\ref{eq:perturbed}) which is defined as
\begin{align}
\Lambda_F = \UU\,\HH_F\, \UU^\dagger = \HH_F^{(0)}+W_F
\end{align}
where we choose $\UU=e^S$ such that $[\Lambda_F,\HH_F^{(0)}] = 0$. This commutation relation is fulfilled if $[\UU,\HH_F^{(0)}]=0$. 
We define the nested commutator as  $[S,\HH_F^{(0)}]_m = [S,[S,\HH_F^{(0)}]_{m-1}]$ with $[S,\HH_F^{(0)}]_0 = \HH_F^{(0)}$.  In this notation the Baker-Campbell-Hausdorff formula takes the form  
\begin{align}
\Lambda_F = \sum_{m=0}^{\infty} \frac{[S,\HH_F]_m}{m!} = \HH^{(0)}_F + \sum_{n=1}^\infty \varepsilon^n \Lambda_F^{(n)}
\end{align}
where we identify
\begin{align}
W_F = \sum_{n=1}^\infty \varepsilon^n \Lambda_F^{(n)} = \sum_{m=1}^{\infty} \frac{[S,\HH_F]_m}{m!}
\label{lambda_expand}
\end{align} 
This is a important equation, because it connects the Hamiltonian $\HH_F$ and the operator $S$ with the perturbation series. We now proceed with  
inserting the expansion
\begin{align}
	S = \sum_{l=1}^{\infty} \varepsilon^l S^{(l)}
\end{align}
as well as Eq. (\ref{eq:perturbed}) in Eq. (\ref{lambda_expand}), which leads to
\begin{align}
\sum_{n=1}^\infty \varepsilon^n \Lambda_F^{(n)}
&= \sum_{m=1}^{\infty} \frac{[\sum_{l=1}^{\infty} \varepsilon^l S^{(l)},\HH_F^{(0)} + \varepsilon\, \HH_F^{(1)}]_m}{m!}
\notag\\
&= \sum_{m=1}^{\infty} \frac{[\sum_{l=1}^{\infty} \varepsilon^l S^{(l)},\HH_F^{(0)}]_m + [\sum_{l=1}^{\infty}\varepsilon^l S^{(l)},\varepsilon\, \HH_F^{(1)}]_m}{m!}
\notag\\
&= \sum_{m=1}^{\infty} \frac{[\sum_{l=1}^{\infty} \varepsilon^l S^{(l)},\HH_F^{(0)}]_m}{m!} +  \sum_{m=1}^{\infty} \frac{ [\sum_{l=1}^{\infty}\varepsilon^l S^{(l)},\varepsilon\, \HH_F^{(1)}]_m}{m!}
\notag\\
&= \sum_{l=1}^{\infty} \varepsilon^l [S^{(l)},\HH_F^{(0)}] +
\sum_{m=2}^{\infty} \frac{[\sum_{l=1}^{\infty} \varepsilon^l S^{(l)},\HH_F^{(0)}]_m}{m!} 
+ \sum_{m=1}^{\infty} \frac{ [\sum_{l=1}^{\infty}\varepsilon^l S^{(l)},\varepsilon\, \HH_F^{(1)}]_m}{m!}
\end{align}
As a result, we arrive at the equation
\begin{align}
[S^{(l)},\HH_F^{(0)}] = \Lambda_F^{(l)} - \Phi^{(l)}_F
\label{eq:commutator_eq}
\end{align} 
with
\begin{align}
\sum_{j=1}^{\infty} \varepsilon^j\Phi^{(j)}_F  = \sum_{m=2}^{\infty} \frac{[\sum_{l=1}^{\infty} \varepsilon^l S^{(l)},\HH_F^{(0)}]_m}{m!} 
+ \sum_{m=1}^{\infty} \frac{ [\sum_{l=1}^{\infty}\varepsilon^l S^{(l)},\varepsilon\, \HH_F^{(1)}]_m}{m!}
\label{eq:Phi}
\end{align}
Utilizing mathematical induction over $m$ we can prove following identities
\begin{align}
\sum_{m=1}^{\infty}[\sum_{l=1}^{\infty} \varepsilon^l S^{(l)},\varepsilon \HH_F^{(1)}]_m
&= \sum_{m=1}^{\infty}   \sum_{n_1=1}^{\infty}\sum_{n_2=1}^{\infty}\cdots \sum_{n_m=1}^{\infty} \,\varepsilon^{1+n_1+n_2+\cdots+ n_m}[S^{(n_1)},[S^{(n_2)},\cdots[S^{(n_m)},\HH^{(1)}_F]\cdots]]\\[0.2cm]
\sum_{m=2}^{\infty}[\sum_{l=1}^{\infty} \varepsilon^l S^{(l)},\HH_F^{(0)}]_m
&= \sum_{m=2}^{\infty}\sum_{n_1=1}^{\infty}\sum_{n_2=1}^{\infty}\sum_{n_3=1}^{\infty}\cdots \sum_{n_m=1}^{\infty} \varepsilon^{n_1+n_2+\cdots+n_m} [S^{(n_1)},[S^{(n_2)},\cdots[S^{(n_m)},\HH^{(0)}_F]\cdots]]
\label{eq:identities}
\end{align}
which lead to a more explicit form of $\sum_{j=1}^{\infty} \varepsilon^j\Phi^{(j)}_F$
\begin{align}
\sum_{j=1}^{\infty} \varepsilon^j\Phi^{(j)}_F &= \sum_{m=2}^{\infty}\sum_{n_1=1}^{\infty}\sum_{n_2=1}^{\infty}\cdots \sum_{n_m=1}^{\infty} \frac{\varepsilon^{n_1+n_2+\cdots+n_m}}{m!}[S^{(n_1)},[S^{(n_2)},\cdots[S^{(n_m)},\HH^{(0)}_F]\cdots]]
\notag\\
&\hspace{0.5cm}+
\sum_{m=1}^{\infty}   \sum_{n_1=1}^{\infty}\sum_{n_2=1}^{\infty}\cdots \sum_{n_m=1}^{\infty} \,\frac{\varepsilon^{1+n_1+n_2+\cdots+n_m}}{m!}[S^{(n_1)},[S^{(n_2)},\cdots[S^{(n_m)},\HH^{(1)}_F]\cdots]]
\label{eq:Phi_induction}
\end{align}
Although this expression is quite complicated, it allows relatively easy and efficient calculation of $\Phi^{(l)}_F$.
To find a solution of Eq. (\ref{eq:commutator_eq}) for $S^{(l)}$ we
adapt the approach of Primas \cite{Primas:1961vi,Primas:1963tg} and use projection operator defined as
\begin{align}
&\Pi(X) = \int P(u)XP(u)\, \dd u &\HH^{(0)}_F = \int\limits_{-\infty}^{\infty} \kappa(u) P(u) \,\dd u 
\label{eq:projection}
\end{align}
By applying the projection operator $\Pi(X)$ on Eq. (\ref{eq:commutator_eq}) we obtain
\begin{align}
\Pi([S^{(l)},\HH_F^{(0)}]) = \Pi(\Lambda_F^{(l)} - \Phi^{(l)}_F) 
\end{align}
Since $[\Lambda_F,\HH_F^{(0)}]=0$, hence $[\UU,\HH_F^{(0)}]=0$, we have
\begin{align}
0=\Pi([S^{(l)},\HH_F^{(0)}]) = \Pi(\Lambda_F^{(l)} - \Phi^{(l)}_F) = \Lambda_F^{(l)} - \Pi(\Phi^{(l)}_F) \implies \Lambda_F^{(l)} = \Pi(\Phi^{(l)}_F)
\label{eq:Lambda}
\end{align}
By introducing the commutation operator
\begin{align}
\Gamma_F (X) = [X, \HH_F^{(0)}],
\hspace{1cm}
\Gamma_F^{-1} (X) = \int \int \frac{P(u)XP(v)}{\kappa(u)-\kappa(v)} \,\dd u \, \dd v
\label{eq:Gamma}
\end{align}
we can rewrite Eq. (\ref{eq:commutator_eq}) as
\begin{align}
\Gamma_F(S^{(l)})= \Pi(\Phi^{(l)}_F) - \Phi^{(l)}_F
\end{align}
and finally arrive at the formal solution 
\begin{align}
S^{(l)} = \Gamma_F^{-1}(\Pi(\Phi^{(l)}_F) - \Phi^{(l)}_F)
\label{eq:S}
\end{align}
The solution has the same form as found by Primas
\cite{Primas:1961vi, Primas:1963tg}, but with differ in the operators $\Pi(X)$ and $\Gamma_F(X)$ given in Eq.(\ref{eq:projection}) and Eq. (\ref{eq:Gamma}), respectively. Ultimately we have all the ingredients to derive each term of the perturbation series.
\subsection{Equivalence of the representations of the Floquet Hamiltonian
  (Eq. (\ref{F}) and Eq. (\ref{Flo}) of the main text)}
The following shows that Eq. (\ref{F}) and Eq. (\ref{Flo}) of the main text is equivalent:
\begin{align}
	\bra{\chi,\mu} \mathcal{H} \ket{\xi, \nu} &= \bra{\chi, \mu} \int D(\Omega)\otimes\widehat{\HH}(\Omega) \, \dd \Omega + \hat{\Omega} \otimes \mathbf{1}\, \ket{\xi,\nu} \notag\\
	& =  \bra{\chi, \mu} \int D(\Omega)\otimes\widehat{\HH}(\Omega)\ket{\xi,\nu}  \dd \Omega +  \bra{\chi, \mu} \hat{\Omega} \otimes \mathbf{1}\,\ket{\xi,\nu} \notag\\
	& = \int  \bra{\chi} D(\Omega)\ket{\xi} \otimes \bra{\mu}\widehat{\HH}(\Omega)\ket{\nu} \dd \Omega +  \bra{\chi} \hat{\Omega} \ket{\xi} \otimes \bra{ \mu}\mathbf{1}\ket{\nu} \notag\\
	& = \int  \braket{\chi|\xi+\Omega} \otimes \bra{\mu}\widehat{\HH}(\Omega)\ket{\nu} \dd \Omega +  \xi \braket{\chi|\xi} \otimes \braket{ \mu|\nu} \notag\\
	& = \int  \delta(\chi-\xi-\Omega) \otimes \widehat{\HH}_{\mu \nu}(\Omega) \,\dd \Omega +  \xi\, \delta(\chi-\xi) \otimes \delta_{\mu \nu} \notag\\
	& = \mathbf{1}\otimes\widehat{\HH}_{\mu \nu}(\chi-\xi) +  \xi\, \delta(\chi-\xi) \otimes \delta_{\mu \nu}
\end{align}
\subsection{Derivation of the first and second-order effective Hamiltonian for a single-mode problem}\label{sec:effective_Hamiltonian}
Next we are going to use the previous results to calculate the first and second-order effective Hamiltonian. For the sake of simplicity, we just present the single mode case, because it already captures the whole procedure. A generalization to the $n$-modal case in retrospect is straight forward, therefore we will just state the results afterwards. The goal is to derive the first three terms of the perturbation series, i.e. 
$\bra{\Omega}\HH^{(0)}_F + \Lambda^{(1)}_F + \Lambda^{(2)}_F\ket{\Omega'}$.  In essence, the only equations we need are Eq. (\ref{eq:Phi_induction}) (or Eq.(\ref{eq:Phi})), Eq. (\ref{eq:Lambda}) and Eq. (\ref{eq:S}). 
For the single mode case the Hamiltonians are
\begin{align}
	\HH^{(0)}_F &= \widehat{\Omega}\\ 
	\HH^{(1)}_F & = \int \widehat{D}(\Omega)\otimes\widehat{\HH}(\Omega)\, \dd \Omega= \Phi^{(1)}_F
\end{align}
In the subsequent derivation the projection operator will be often applied on the translation operator, which results in a delta distribution
\begin{align}
\Pi(\widehat{D}(\Omega)) 
 = \delta(\Omega)
\end{align}
The projection of $\Phi^{(1)}_F$  simply leads to
\begin{align}
\Lambda_F^{(1)}=\Pi\left(\Phi^{(1)}_F\right) 
=\Pi\left(\int \widehat{D}(\Omega)\otimes\widehat{\HH}(\Omega)\, \dd \Omega\right)
= \int \delta(\Omega)\otimes\widehat{\HH}(\Omega)\, \dd \Omega = \widehat{D}(0)\otimes \widehat{\HH}(0)
\end{align}
Inserting it into the Eq. (\ref{eq:S}) gives
\begin{align}
S^{(1)} = \Gamma^{-1}\left(\Pi\left(\Phi^{(1)}_F\right)-\Phi^{(1)}_F\right) 
=\Gamma^{-1}\left( \widehat{D}(0)\otimes \widehat{\HH}(0)-\int \widehat{D}(\Omega)\otimes\widehat{\HH}(\Omega)\, \dd \Omega\right)
= -\, PV \int\frac{
	\widehat{D}(\Omega)\otimes\widehat{\HH}(\Omega)\,  }{\Omega} \,\dd \Omega
\end{align}
Notice that we use Cauchy principal value (PV) for the regularization of the integral defined as
\begin{align}
PV \int_{a}^c f(x) = \lim \limits_{\epsilon \to 0^{+}} \left[\int_{a}^{b-\epsilon} f(x) \,dx + \int_{b+\epsilon}^{c} f(x) \, dx\right]
\end{align}
Using this regularization technique is justified, since we subtract $\widehat{D}(0) \otimes \widehat{\HH}(0)$, which is the kernel of the integral at the critical value $\Omega = 0$ and therefore can be exclude it from the integration. 
Equipped with the expression for $S^{(1)}$ we move to the second order 
\begin{align}
\Phi^{(2)}_F &= [S^{(1)},\HH^{(1)}_F]+\frac{1}{2} [S^{(1)},[S^{(1)},\HH_F^{(0)}]\\
             &= - \,PV \int \frac{[\widehat{\HH}(\Omega),\HH(0)]\widehat{D}(\Omega)}{\Omega}\,\dd \Omega
               +\frac{1}{2} \, PV
\int \int
 \frac{[\widehat{\HH}(\Omega),\widehat{\HH}(\Omega')]\widehat{D}(\Omega+\Omega')}{\Omega}\,\dd \Omega \, \dd \Omega'
\end{align}
which results in
\begin{align}
\Lambda_F^{(2)}=\Pi\left(\Phi^{(2)}_F\right)= \frac{1}{2}\, PV
\int
\frac{[\widehat{\HH}(\Omega),\widehat{\HH}(-\Omega)]\widehat{D}(0)}{\Omega}\,\dd \Omega
\end{align}
Finally we express the Hamiltonian in the product basis
\begin{align}
\bar{\HH}^{(0)} + \bar{\HH}^{(1)}  + \bar{\HH}^{(2)}
=  \bra{\Omega,\mu} \HH_{F}^{(0)} + \Lambda^{(1)}_F + \Lambda^{(2)}_F \ket{\Omega',\nu}
\end{align}
with
\begin{align}
\bar{\HH}^{(0)} = \widehat{\Omega}    \hspace{1cm}
\bar{\HH}^{(1)} =  \int\limits_{-\infty}^{\infty}  \widehat{\HH}(\Omega)\, \dd \Omega  \hspace{1cm}
\bar{\HH}^{(2)} = \frac{1}{2} \, PV
\int \int 
\frac{[\widehat{\HH}(\Omega)
	,\widehat{\HH}(-\Omega)]}
{\Omega}
\dd\Omega
\end{align}
As expected this results are similar to the corresponding equations known from Floquet theory. The main difference is that we have an integral and a translation operator instead of a sum and a ladder operator.\\
One of the main differences in the higher modal cases is the appearance of resonance conditions between the modes. For the n-modal case the resonance conditions are
\begin{align}
\Omega^{(0)}_1 + \Omega^{(0)}_2 + \cdots + \Omega^{(0)}_n = 0
\end{align}
The derivation of the effective Hamiltonians is quite similar and leads to the first and second-order effective Hamiltonian, for the bimodal case 
\begin{align}
\bar{\HH}^{(1)} &=  \int\limits_{-\infty}^{\infty} \int\limits_{-\infty}^{\infty}  \widehat{\HH}(\Omega_1,\Omega_2)\, \dd \Omega_1\, \dd \Omega_2
\end{align}
\begin{align}
\bar{\HH}^{(2)} &= 
\int \dd\Omega_1 
\int \dd\Omega_1^{(0)}
\int \dd\Omega_2
\int \dd\Omega_2^{(0)}
\frac{[\widehat{\HH}(\Omega_1,\Omega_2)
	,\widehat{\HH}(\Omega_1^{(0)}-\Omega_1,\Omega_2^{(0)}-\Omega_2)]}
{\Omega_1 + \Omega_2}
\end{align}

\section{Calculation of the frequency-domain interaction-frame trajectory}
\label{sec:interaction-frame-trajectory}
\subsection{Propagator}
\label{subsec:propagator}
The propagator of the $j$th time slice $\UU_j$ (Eq. (\ref{eq:single_rot})), has the complex $2\times 2$ matrix representation
\begin{align}
\UU_j 
&= \exp\left(-i\,\HH_j \delta t\right) = \frac{1}{|\theta|\delta t} \left(
\begin{array}{cc}
|\theta| \cos \left(\frac{|\theta|\delta t}{2}\right)+i\,\theta_{z} \sin
\left(\frac{|\theta|\delta t}{2}\right) & \sin \left(\frac{|\theta|\delta t}{2}\right)
(\theta_y+i\,\theta_x) \\
\sin \left(\frac{|\theta|\delta t}{2}\right) (-\theta_y+i\,\theta_x) &
|\theta| \cos \left(\frac{|\theta|\delta t}{2}\right)-i\,\theta_z \sin
\left(\frac{|\theta|\delta t}{2}\right)\\
\end{array}
\right)
\label{generalU}
\end{align}
where we dropped the index $j$ of $\theta$ for sake of simplicity.
This representation is used for derivation of Eq. (61) in the main text.
\subsection{The frequency-domain interaction-frame trajectory for a cyclic
  pulse scheme}\label{subsec:cyclic trajectory}
The interaction-frame trajectory of a cyclic pulse scheme with an
basic element of duration $\tau$ and an overall duration $N\tau$ can be
written as
\begin{align}
    a_{\mu\nu}(t) = 
    \begin{dcases}
        a_{\mu\nu}(t) & 0 \leq t \leq \tau\\
        \sum_{\chi}a_{\mu\chi}(t-\tau)a_{\chi\nu}(\tau) & \tau \leq t \leq 2\tau\\
        \sum_{\chi}a_{\mu\chi}(t-2\tau)a_{\chi\nu}(2\tau) & 2\tau \leq t \leq 3\tau\\
        \hspace{2cm}\vdots & \quad\quad\vdots\\
        \sum_{\chi}a_{\mu\chi}(t-(N-1)\tau)a_{\chi\nu}((N-1)\tau) & (N-1)\tau \leq t \leq N\tau
    \end{dcases}
\end{align}
Fourier transformation leads to
\begin{align}
  \widehat{a}_{\mu\nu}(\Omega) &= \frac{1}{2\pi N \tau} \biggl(\sum_{\chi}
  \int_0^\tau a_{\mu\nu}(t)e^{-i \Omega t} dt
  +\int_\tau^{2\tau} a_{\mu\chi}(t-\tau)a_{\chi\nu}(\tau)e^{-i \Omega t} dt
  +\int_{2\tau}^{3\tau} a_{\mu\chi}(t-2\tau) a_{\chi\nu}(2\tau)e^{-i \Omega t} dt \\
  &\quad\quad + \cdots + \int_{(N-1)\tau}^{N\tau} a_{\mu\chi}(t-(N-1)\tau) a_{\chi\nu}((N-1)\tau)e^{-i \Omega t} dt \biggr)
\notag\\
  &= \frac{1}{2\pi N \tau} \sum_{\chi} \sum_{n=1}^{N}
     \int_{(n-1)\tau}^{n\tau} a_{\mu\chi}(t-(n-1)\tau) a_{\chi\nu}((n-1)\tau)e^{-i \Omega t} dt\\
  &=\frac{1}{2\pi N \tau} \sum_{\chi} \sum_{n=1}^{N} e^{-i\Omega(n-1)\tau}
    \int_{0}^{\tau} a_{\mu\chi}(t') a_{\chi\nu}((n-1)\tau)e^{-i \Omega t'} dt'\\
  &=\frac{1}{2\pi N \tau} \sum_{\chi}
    \int_{0}^{\tau} a_{\mu\chi}(t')e^{-i \Omega t'} dt'  \sum_{n=0}^{N-1} e^{-i\Omega n \tau} a_{\chi\nu}( n \tau)\\
  &=      \sum_\chi \bar{a}_{\mu\chi}(\Omega)\, g^{(N)}_{\chi \nu}(\Omega)    
\end{align}
and therefore
\begin{align}
    g^{(N)}_{\mu \nu}(\Omega) \coloneqq \frac{1}{N}\sum_{n=0}^{N-1} e^{-i n \Omega \tau} a_{\mu \nu}(n\tau)
\end{align}
Notice that $a_{\mu\nu}(n\tau) = [\mathbf{a}^n(\tau)]_{\mu\nu}$ where
$\mathbf{a}(\tau)$ is the matrix with the elements
$a_{\mu\nu}(\tau)$.
Furthermore we have 
\begin{align}
    \bar{a}_{\mu\nu}(\Omega) &= \frac{1}{2\pi \tau} \int\limits_{0}^{\tau} {a}_{\mu\nu}(t)e^{-i\Omega t} dt
     \\
     &= \Fou\{\widetilde{a}_{\mu\nu}(t) \Pi\left(t/\tau-1/2\right)\}
     \\
     &=\Fou\{\widetilde{a}_{\mu\nu}(t)\}*\mathcal{F}\left\{ \Pi(t/\tau-1/2) \right\}
\end{align}
and
\begin{align}
  \Fou\{\widetilde{a}_{\mu\nu}(t)\}=\widetilde{a}_{\mu\nu}(\Omega) = \frac{1}{2\pi} \int\limits_{-\infty}^{\infty} {a}_{\mu\nu}(t\,\textrm{mod}\tau)e^{-i\Omega t} dt
\end{align}
Notice that $\widetilde{a}_{\mu\nu}(\Omega)$ is just calculated using FFT.
As described in the main text, the special case of a cyclic
interaction-frame trajectory occurs when  $\mathbf{a}(\tau)=\mathbf{1}$.
\end{widetext}
\end{document}